\documentclass{lmcs}
\pdfoutput=1

\usepackage[utf8]{inputenc}

\usepackage{lastpage}
\lmcsdoi{17}{1}{15}
\lmcsheading{}{\pageref{LastPage}}{}{}%
{Sep.~23,~2019}{Feb.~12,~2021}{}


\usepackage{graphicx}
\usepackage{bussproofs}
\usepackage{amsmath}
\usepackage{amssymb}
\usepackage{bbm}
\usepackage{stmaryrd}
\usepackage{listings}

\lstdefinestyle{numbered}{
  numbers=left,
  xleftmargin=7.5mm, 
}

\begin{document}
\title{Output-sensitive Information flow  analysis\rsuper*}
\titlecomment{{\lsuper*}This work is supported by the French National Research Agency in the framework of the ``Investissements d' avenir'' program (ANR-15-IDEX-02)}

%
%


\author[C.~Ene]{Cristian Ene}  
\address{Univ. Grenoble Alpes, CNRS, Grenoble INP
, VERIMAG, 38000 Grenoble, France}

\author[L.~Mounier]{Laurent Mounier}    
\email{Cristian.Ene@univ-grenoble-alpes.fr}  
\email{Laurent.Mounier@univ-grenoble-alpes.fr}  
\email{Marie-Laure.Potet@univ-grenoble-alpes.fr}  

\author[ML.~Potet]{Marie-Laure Potet}  

%
%
%
%
\begin{abstract}
{\it Constant-time} programming is a countermeasure  to prevent cache based attacks where programs should not perform memory accesses that depend on secrets. In some cases this policy can be safely relaxed if one can prove that the program does not leak more information than the public outputs of the computation. 

We propose a novel approach for verifying constant-time programming based on a new information flow property, called {\it output-sensitive noninterference}. 
Noninterference states that a public observer cannot learn anything about the private data. Since real systems need to intentionally declassify some information, this property is too strong in practice. 
In order to take into account public outputs we proceed as follows: instead of using complex explicit declassification policies, we partition variables in three sets: input, output and leakage variables. Then, we propose a typing system to statically check that leakage variables do not leak {\em more information about the secret inputs than the public normal output}. The novelty of our approach is that we track the dependence of leakage variables with respect not only to the initial values of input variables (as in classical approaches for noninterference), but taking also into account the final values of output variables. We adapted this approach to LLVM IR and we developed a prototype to verify 
LLVM implementations.
\end{abstract}

\maketitle              

\newcommand\todo[1]{{\color{red} {\bf #1}}}
\newcommand\comm[1]{}
\newcommand\Dist{\textsc{Dist}}
\newcommand\cO{\mathcal{O}}
\newcommand\cL{\mathcal{L}}
\newcommand\bF{\textbf{F}}
\newcommand\bT{\textbf{T}}
\newcommand\mbs{\mathbf{s}}
\newcommand\ndec{\ensuremath{\not \succ}}
\newcommand\sem[1]{\ensuremath{[\![#1]\!]}}
\newcommand\ssem[1]{[\![#1]\!]}
\newcommand\wfn[1]{\ensuremath{\langle #1 \rangle_{\not\succ}}}
\newcommand\fir[1]{\ensuremath{\langle #1 \rangle_{\cong}}}
\newcommand\cP{\mathcal{P}}
\newcommand\cR{\mathcal{R}}
\newcommand\cS{\mathcal{S}}
\newcommand\cT{\mathcal{T}}
\newcommand\cI{\mathcal{I}}
\newcommand\nat{\mathbbm{N}}
\newcommand\Adv{\textsf{Adv}}

\newcommand\HASH{\ensuremath{\mathcal{H}}}
\newcommand\cH{\ensuremath{\mathcal{H}}}
\newcommand\randd[2]{{#1}\stackrel{r}{\leftarrow}{#2}}
\newcommand\cA{\mathcal{A}}
\newcommand\cB{\mathcal{B}}
\newcommand\cG{\mathcal{G}}
\newcommand\pr{\textsf{Pr}}
\newcommand\mina{\mathbf{atomO}}
\newcommand\wf[1]{\ensuremath{\mathcal{WF}(#1)}}
\newcommand\ac[1]{\ensuremath{\mathcal{WF}(#1)}}
\newcommand\gt[2]{\ensuremath{\mathcal{GT}_{#1}(#2)}}

\newcommand\evar{\ensuremath{Var^{e}}}

\newcommand\deff{\stackrel{def}=}

\newcommand\trH[3]{\ensuremath{#1\{#2\}#3}}

\newcommand\mapr[3]{\ensuremath{#1=_{#3}#2}}



\newcommand\iifthenelse[3]{\textsf{If } #1 \textsf{ then } #2 \textsf{
    else } #3 \textsf{ fi }}
\newcommand\whilep[2]{\textsf{While } #1 \textsf{ Do } #2 \textsf{
    oD }}
\newcommand\skipp{\textsf{skip}}

\newcommand\sbs{\sqsubseteq}
\newcommand\ubs{\sqsupseteq}
\newcommand\js{\sqcup}
\newcommand\ms{\sqcap}
\newcommand\exec[3]{\langle #1, #2 \rangle \Longrightarrow  #3}

\newcommand\type[3]{#1\vdash  #2 : #3}
\newcommand\typeg[2]{$\type{\Gamma}{#1}{#2}$}
\newcommand\nntype[2]{$#1\vdash_{\alpha} #2$}
\newcommand\ntype[2]{$\vdash_{\alpha} #2$}
\newcommand\ctype[2]{$#1\vdash_{\alpha}^{ct} #2$}
\newcommand\stype[2]{\ensuremath{#1} $\vdash_{S}$ #2}
\newcommand\ptype[2]{\ensuremath{#1} $\vdash #2$}
\newcommand\ntypeo[2]{$#1\vdash #2$}
\newcommand\ntypeg[2]{\type{\gamma}{#1}{#2}}
\newcommand\rulet[2]{\frac{#1}{#2}}
\newcommand\rand[2]{{#1}\stackrel{r}{\leftarrow}{#2}}
\newcommand\addT[3]{\ensuremath{#1[#2] \uplus #3}}

\newcommand\red[2]{\stackrel{#1}{#2}}
\newcommand\reda[1]{\stackrel{#1}{\longrightarrow}}
\newcommand\rd[1]{\textbf{r}(#1)}
\newcommand\wt[1]{\textbf{w}(#1)}
\newcommand\ju[1]{\textbf{j}(#1)}
\newcommand\br[1]{\textbf{b}(#1)}
\newcommand\vc{\overrightarrow}
\newcommand\aff{\mathbf{def}}
\newcommand\affi{\mathbf{def^I}}
\newcommand\affo{\mathbf{def^O}}
\newcommand\ba[1]{\overline{#1}}
\newcommand\di{\diamond}

\newcommand\repl[2]{#1 \di #2}
\newcommand\replga{\repl{\Gamma}{\alpha}}

\newcommand\While{\textsf{While}}

\newcommand\omi{\omega}
\newcommand\gami{\gamma}

\newcommand\bi{b_{init}}
\newcommand\be{b_{end}}

\section{Introduction}
\label{sec:intro}

An important task of cryptographic research is to verify cryptographic
implementations for security flaws, in particular to avoid so-called timing attacks. 
Such attacks consist in measuring the execution time 
of an implementation on its execution platform. For instance,
Brumley and Boneh \cite{bru2005}
showed that it was possible to mount remote timing attacks by
against OpenSSL's implementation of the RSA decryption operation 
and to recover the key. Albrecht and Paterson
\cite{alb2016} showed that the two levels of protection offered against
the Lucky 13 attack from \cite{alb2013} in the first release of
the new implementation of TLS were imperfect.
A related class of attacks are {\em cache-based attacks} 
in which a malicious party is able to obtain
memory-access addresses of the target program which may depend on
secret data through observing cache accesses. Such attacks
allow to recover the complete AES keys \cite{gul2011}.

A possible countermeasure is to follow a very
strict programming discipline called {\bf constant-time programming}.
Its principle is to avoid branchings {\bf controlled} by secret data 
and memory load/store operations {\bf indexed} by secret data. 
Recent secure C libraries such as NaCl \cite{ber2012} or
mbedTLS\footnote{mbed TLS (formerly known as
  PolarSSL). https://tls.mbed.org/} follow this programming
discipline. Until recently, there was no rigorous proof that constant-time
algorithms are protected against cache-based attacks.  Moreover, many
cryptographic implementations such as PolarSSL AES, DES, and RC4 make
array accesses that depend on secret keys and are not constant time.
Recent works \cite{bar2014,almeida2016,bla2017} fill this gap and
develop the first formal analyzes that allow to verify if programs are
correct with respect to the constant-time paradigm.

An interesting extension was brought by Almeida et al.~\cite{almeida2016} 
who enriched the constant-time paradigm \emph{``distinguishing not only between public and private input values, but
also between private and publicly observable output values}''. 
This distinction raises interesting technical and
theoretical challenges.  Indeed, constant-time implementations in
cryptographic libraries like OpenSSL include optimizations for which
paths and addresses can depend not only on public input values, but
also on publicly observable output values. Hence, considering only input
values as non-secret information would thus incorrectly characterize
those implementations as non-constant-time. \cite{almeida2016} also develops a
verification technique based on {\sl symbolic execution}. 
However, the soundness of their approach depends in practice on the soundness of the underlying symbolic execution
engine, which is very difficult to guarantee for real-world programs with loops. 
Moreover, their product construction can be very expensive in the worst case.

In this paper we deal with {\sl statically checking programs} for {\bf
  output-sensitive constant-time} correctness: programs can still do
branchings or memory accesses controlled by secret data if the
information that is leaked is subsumed by the normal output of the
program. 
To give more intuition about the property that we want to deal with,
let us consider the following example, where $ct\_eq$ is a constant
time function that allows to compare the arguments:
\begin{lstlisting}[language=C]
good = 1;
i = 0;
for ( ; i<B_Size; i++){
    good = good & ct_eq(secret[i],in_p[i]);
}
if (!good){ 
    i = 0;
    for( ; i<B_Size; i++) secret[i] = 0; 
}
return good;
\end{lstlisting}
Let suppose that the array variable $secret$ is secret, and all the
other variables are public.  Intuitively this a sort of one-time check
password verifying that $in\_p=secret$ and otherwise
overwrites the array $secret$ with zero.
Obviously, this function is not constant-time as the variable
$good$ depends on $secret$, and hence branching on $good$ violates 
the principles of constant-time programming.
It is easy to transform this program into an equivalent one which is
constant time. For example one could replace
\begin{lstlisting}[language=C]
if (!good){ 
   i = 0;
   for( ; i<B_Size; i++) secret[i] = 0; 
}
\end{lstlisting}
by
\begin{lstlisting}[language=C]
i = 0;
for ( ; i<B_Size; i++){
   secret[i] = secret[i] & ct_eq(good,1);
}
\end{lstlisting}
But branching on $good$ is a benign optimization, since
anyway, the value of $good$ is the normal output of the
program.  Hence, even if the function is not constant-time, it should
be considered {\bf output-sensitive constant time} with respect to its
specification. Such optimization opportunities arise whenever the
interface of the target application specifies what are the publicly
observable outputs, and this information is sufficient to classify the
extra leakage as benign \cite{almeida2016}.

The objective of this work is to propose a {\sl static method} to
check if a program is {\bf output-sensitive side-channel secure}.
We emphasize that our goal is {\bf not} to verify that the legal 
 output leaks ``too much'', but rather to ensure that the unintended 
(side-channel) output does not leak {\bf more than} this legal output.

First, we propose a novel approach for verifying \emph{constant-time like
security} based on a new information flow property, called {\it
  output-sensitive noninterference}. Information-flow security
prevents confidential information to be leaked to public
channels. Noninterference states that a public observer cannot learn
anything about the private data. Since real systems need to
intentionally declassify some information, this property is too strong.
A possible alternative is {\it relaxed noninterference} which
allows to specify explicit {\it downgrading policies}.
For instance, in~\cite{sab2003}, the authors proposed the ``delimited release security property'' that captures the leakage by a  ``declassify'' primitive, which allows to state \emph{explicitly} what is permitted to be leaked with respect to the \emph{initial state}. 

In order to take into account public outputs
while staying independent of how programs intentionally declassify information, 
we develop an alternative solution: instead of
using complex explicit policies for functions, we partition variables
in three sets: input, output and \textit{leakage variables}. Hence we
distinguish between the legal public output and the information that
can leak through side-channels, expressed by adding fresh additional leakage variables.  
Then we propose a typing system that can statically check that leakage variables do not 
leak more secret information than the {\bf public normal output}. 
Therefore, in our case, the legal leakage is \emph{implicitly} defined by the normal output of the program
and not by any dedicated language primitive.  
The novelty of our approach is that we track the dependence of leakage
variables with respect to both the {\em initial value of input variables} 
(as  classically the case for noninterference) and {\em final values of output variables},   
which hardens the analysis.

As an application of this new non-interference property, we show how to verify that a program written in a high-level
language is output-sensitive constant time secure by using this typing system. 

Since timed and cache-based attacks target the executions of programs,
it is important to carry out this verification in a language 
close to the machine-executed assembly code. Hence, we adapt our approach to a generic unstructured assembly
language inspired from LLVM and we show how we can verify programs
coded in LLVM. Finally, we developed a prototype tool implementing our type system 
and we show how it can be used to verify LLVM implementations.

To summarize, this work makes the following contributions described above:
\begin{itemize}
\item in section \ref{s:2} we reformulate output-sensitive constant-time 
  as a new interesting noninterference property and we provide a sound
  type system that guarantees that well-typed programs are output-sensitive noninterferent; \\
\item in section \ref{s:3} we show that this general approach can be
  used to verify that programs written in a high-level language are
  output-sensitive constant time; \\
\item in section \ref{llvm:sec} we adapt our approach to the LLVM-IR
  language and we develop a prototype tool that can be used to  verify LLVM
  implementations.
\end{itemize}


\section{Output-sensitive non-interference}\label{s:2}

In this section we introduce {\it output-sensitive noninterference} as a new information flow property 
that aims to extend classical noninterference definitions in two directions:
\begin{itemize}
\item this property should be able to characterize \emph{general side channel attacks};
\item in the same time, it should be accurate enough to capture the precise side channel information leaked \emph{beyond} the regular output of the program.
\end{itemize}

In order to take into account public outputs
while staying independent of how programs intentionally declassify information,
we partition variables in three sets: input, output and \textit{leakage variables}. Hence we
distinguish between the legal public output and the information that
can leak through side-channels, expressed by adding fresh additional leakage variables.
These leakage variables are updated each time that some information is leaked to the environment.
For instance, for some leakage variable $x_l$, 
if an adversary can get some knowledge about the branch that will be taken in case of a test condition $if~e~then~c_1~else~c_2$, 
we update $x_l$ with a dependency on $e$. 
Similarly, for some assignment $x~:=~f(e_1, \dots ,e_n)$, if the evaluation of $f$ is time-dependent on its arguments, then we update 
$x_l$ with dependencies on $e_1, \dots ,e_n$. In this paper we focus on constant-time security, but our approach could be used as well 
for other side-channel attacks. 

In order to precisely characterize the side channel information leaked additionally with respect to the normal output of the program,
we also consider \textit{output variables}.

The novelty of our approach is that we track the dependency of leakage
variables with respect to both the {\em initial value of input variables}
(as  classically the case for noninterference) and {\em final values of output variables},
which hardens the analysis.

The rest of the section is structured as follows: first we introduce the {\it While} language and we formulate the definition of output-sensitive noninterference.
Then we propose a typing system that can statically check that leakage variables do not
leak more secret information than the public normal output. We conclude the section by proving the soundness of this typing system.


\subsection{The \While \ language and Output-sensitive noninterference}
In order to reason about the security of the code,  we first develop our framework in {\it While}, a  simple high-level structured programming language. In  section \ref{s:3} we shall enrich this simple language with arrays and in section \ref{llvm:sec} we adapt our approach to a generic unstructured assembly language. The syntax of While programs is listed below: 
\[
\begin{array}{lll}
c & ::= & x:=e ~|~ \skipp~|~ c_1;c_2~|~ \iifthenelse{e}{c_1}{c_2}~|~ \whilep{e}{c}
\end{array}
\]
 Meta-variables $x, e$ and $c$ range over the sets of  program variables $Var$, expressions and programs,  respectively. We leave the syntax of expressions unspecified, but we assume they are deterministic and side-effect free. The semantics is shown in Figure \ref{fig:while:sem}.  The reflexive and transitive closure of $\longrightarrow$ is denoted by $\Longrightarrow$.  A state $\sigma$ maps variables to values, and we write $\sigma(e)$ to denote the value of expression $e$ in state $\sigma$. A configuration $(c, \sigma)$ is a program $c$ to be executed along with the current state $\sigma$.  
\begin{figure*}[!tbp]
\[
\begin{array}{ll}

\begin{minipage}[h]{0.5\linewidth}
\begin{prooftree} 
\AxiomC{ }
\UnaryInfC{$(x:=e,\sigma)\longrightarrow \sigma[x\mapsto\sigma(e)]$}
\end{prooftree}
\end{minipage}

&   

\begin{minipage}[h]{0.4\linewidth}
\begin{prooftree} 
\AxiomC{ }
\UnaryInfC{$(\skipp,\sigma)\longrightarrow \sigma$}
\end{prooftree}
\end{minipage}

\\

\\
\begin{minipage}[h]{0.5\linewidth}
\begin{prooftree} 
\AxiomC{$(c_1,\sigma)\longrightarrow \sigma'$}
\UnaryInfC{$( c_1;c_2,\sigma)\longrightarrow (c_2, \sigma')$}
\end{prooftree}
\end{minipage}
&

\begin{minipage}[h]{0.4\linewidth}
\begin{prooftree} 
\AxiomC{$(c_1,\sigma)\longrightarrow (c'_1,\sigma')$}
\UnaryInfC{$( c_1;c_2,\sigma)\longrightarrow (c'_1;c_2, \sigma')$}
\end{prooftree}
\end{minipage}
\\

\\

\begin{minipage}[h]{0.5\linewidth}
\begin{prooftree} 
\AxiomC{$\sigma(e)  = 1 ~ ? ~ i=1 : \ i=2$}
\UnaryInfC{$(\iifthenelse{e}{c_1}{c_2},\sigma)\longrightarrow (c_i, \sigma)$}
\end{prooftree}
\end{minipage}

&

\\

\begin{minipage}[h]{0.5\linewidth}
\begin{prooftree}
\AxiomC{}
	\UnaryInfC{$(\whilep{e}{c},\sigma)\longrightarrow (\iifthenelse{e}{c;\whilep{e}{c}}{\skipp},\sigma)$}
\end{prooftree}
\end{minipage}

\end{array}
\]
\caption{Operational semantics of the \ \While \ language}\label{fig:while:sem}
\end{figure*}
Intuitively, if we want to model the security of some program $c$ with
respect to side-channel attacks, we can assume that there are three
special subsets of variables: $X_I$ the public input variables, $X_O$
the public output variables and $X_L$ the variables that leak
information to some malicious adversary. Then, output sensitive
nonintereference asks that every two complete executions starting with
$X_I$-equivalent states and ending with $X_O$-equivalent final states must
be indistinguishable with respect to the leakage variables $X_L$.
In the next definition, $=_X$ relates two states that coincide on all variables belonging to $X$.

\begin{defi}[adapted from \cite{almeida2016}]\label{iol-sec}
Let $X_I, X_O, X_L\subseteq Var$  be three sets of variables, intended to represent the input, the output and the leakage of a program.  A program $c$ is {\bf  ($X_I, X_O, X_L$)-output-sensitive non-interferent} when all its executions starting with $X_I $-equivalent stores and leading to $X_O$-equivalent final stores,  give $X_L$-equivalent final stores. 
Formally, for all $\sigma, \sigma', \rho, \rho'$, if $\exec{c}{\sigma}{\sigma'}$ and $\exec{c}{\rho}{\rho'}$ and $\sigma =_{X_I} \rho $ and $\sigma'=_{X_O}\rho'$, then $\sigma'=_{X_L}\rho'$.
\end{defi}

\begin{exa}
To illustrate the usefulness of the output-sensitive non-interference property let us consider again the example given in the previous section,
shown on Figure~\ref{fig:c-ex0}.
\label{ex-intro}
\begin{figure*}[!htb]
\begin{center}
\begin{lstlisting}[language=C,style=numbered]
good = 1;
i = 0;
for (; i<B_Size; i++){
   good = good & ct_eq(secret[i],in_p[i]);
}
if (!good) { 
   i = 0;
   for(; i<B_Size; i++) 
      secret[i] = 0; 
}
return good;
\end{lstlisting}
        \caption{C code example from Section~1}
        \label{fig:c-ex0}
\end{center}
\end{figure*}
Then, in order to reduce its output-sensitive constant time security to the output-sensitive non-interference property (as in Definition~\ref{iol-sec}), 
first, we add a fresh variable {\tt xl} to accumulate the leakage information (using the abstract concatenation operator {\tt @}).
Then, variable {\tt xl} is updated with the boolean condition of each branching instruction (i.e., at lines 3, 7, 9, 12 and 16) and with each expression used as an index array (i.e., at lines 5 and 14).
The result of this transformation is given on Figure~\ref{fig:c-ex1} (we denote by {\tt ct\_eq\_ni} the function obtained by applying recursively the same 
transformation to {\tt ct\_eq}).
\begin{figure*}[!htb]
\begin{center}
\begin{lstlisting}[language=C,style=numbered]
good = 1;
i = 0;
xl = xl @ (i<B_Size);
for (; i<B_Size; i++){
   xl = xl @ i;
   good = good & ct_eq_ni(secret[i],in_p[i]);
   xl = xl @ (i<B_Size);
}
xl = xl @ good;
if (!good) { 
    i = 0;
    xl = xl @ (i<B_Size);
    for(; i<B_Size; i++) { 
       xl = xl @ i;
       secret[i] = 0; 
       xl = xl @ (i<B_Size);
    }
}
return good;
\end{lstlisting}
        \caption{C code example from Section~1 with explicit leakage}
        \label{fig:c-ex1}
\end{center}
\end{figure*}
We consider the following sets of variables: $X_I=\{\mbox{{\tt in\_p}}\}$, $X_L=\{\mbox{{\tt xl}}\}$ and $X_O=\{\mbox{{\tt good}}\}$.
One can check that, for each pair of executions starting with the same value of {\tt in\_p} and ending with the same value of {\tt good}, 
then the final value of {\tt xl} will be the same.
Then, according to Definition~\ref{iol-sec}, this program is ($X_I, X_O, X_L$)-output-sensitive non-interferent.
\end{exa}

In Section~\ref{s:3}  we will show that this transformation is general and, moreover, whenever the transformed program satisfies 
Definition~\ref{iol-sec}, then the initial program is output-sensitive constant-time, i.e.,  it can still do
branchings or memory accesses controlled by secret data if the information that is leaked is subsumed by its normal output.
Note that the code of Figure~\ref{fig:c-ex1} is a simplified 
application of this transformation to the code of Figure~\ref{fig:c-ex0}, i.e., we omitted trivial assignments of the form {\tt xl = xl}.

\subsection{Preliminary definitions and notations}

As usual, we consider a flow lattice of security {\bf types} $\cL$ (also called security levels).  An element $x$ of  $\cL$ is an atom if $x\not=\bot$  and there exists no element $y \in \cL$ such that $\bot \sqsubset y \sqsubset x$. A lattice is called {\bf atomistic} if every element \emph{$x \in \cL$ is the join of the set of atoms below it}~\cite{Atl}. This set is denoted by  $At(x)$.
\begin{asm}\label{as1}
Let $(\cL, \ms, \js, \bot, \top)$ be an atomistic bounded lattice. As usual, we denote $t_1\sbs t_2$ iff $t_2 = t_1 \js t_2$.  We assume that there exists a distinguished subset $\cT_O\subseteq \cL$ of atoms.
Hence, from the above assumption, for any $\tau_o, \tau'_o\in \cT_O$ and for any $t_1,t_2\in \cL$:  it holds that 
\begin{itemize}[align=left]
\item[(A1)]\label{a1} $\tau_o\sbs \tau'_o$ implies $\tau_o = \tau'_o$,
\item[(A2)]\label{a2} $\tau_o\sbs t_1 \sqcup t_2$ implies $\tau_o\sbs t_1$ or $\tau_o\sbs t_2$, 
\item[(A3)]\label{a3} $\tau_o\sbs t_1$ implies that there exists $t\in \cL$ such that $t_1= t \sqcup \tau_o$ and $\tau_0 \not\sbs t$.
\end{itemize}
\end{asm}

The assumption that our set $\cL$ of security types is an atomistic lattice provides a general structure which is sufficient for our purposes: it ensures the existence of decompositions in
atoms for any element in the lattice, and the ability to syntactically replace an atomic type by another (not necessarily atomic) type.

A type environment  $\Gamma : Var \mapsto \cL$ describes
the {\bf security levels} of variables and the dependency with respect to the {\bf current} values of variables in $X_O$. 
In order to catch dependencies with respect to current values of output variables, we associate to each output variable $o\in X_O$ a fixed and unique symbolic type $\alpha(o)\in \cT_O$. 
For example if some variable $x\in Var$ has the type $\Gamma(x)=Low \js \alpha(o)$, it means that the value of $x$ depends only on public input and the current value of the output variable $o\in X_O$. 

Hence, we assume that there is a fixed  injective mapping $\alpha : X_0\mapsto \cT_0$ such that $\displaystyle\bigwedge_{o_1,o_2\in X_O} \big(o_1\not=o_2  \Rightarrow \alpha(o_1)\not=\alpha(o_2)\big) \wedge \bigwedge_{o\in X_O} \big(\alpha(o)\in \cT_O \big)$.  
We extend mappings $\Gamma$ and $\alpha$  to sets of variables in the usual way: given $A\subseteq Var$ and $B\subseteq X_O$ we note $\displaystyle\Gamma(A) \deff \bigsqcup_{x\in A} \Gamma(x)$ ,  $\displaystyle\alpha(B) \deff \bigsqcup_{x\in B} \alpha(x)$.

Our type system aims to satisfy the following output sensitive non-interference condition: if
the \emph{final} values of output variables in $X_O$ remain the same, only changes
to \emph{initial} inputs with types $\sbs t$ should be visible to \emph{leakage} outputs with type
$\sbs t \sqcup \alpha(X_O)$. More precisely, given a derivation
\nntype{}{\trH{\Gamma}{c}{\Gamma'}}, the final value of a variable
$x$ with final type $\Gamma'(x) = t \js \alpha(A)$ for some $t\in \cL$
and $A\subseteq X_O$ should depend at most on the initial values of
those variables $y$ with initial types $\Gamma(y)\sbs t$ and on the
final values of variables in $A$. We call ``real dependencies'' the
dependencies with respect to initial values of variables and
``symbolic dependencies'' the dependencies with respect to the current
values of output variables.  Following \cite{Hunt91} we formalize the
non-interference condition satisfied by the typing system using
reflexive and symmetric relations.

We write $=_{A_0}$ for relation which relates mappings which are equal on all values in $A_0$ i.e.  for two mappings $f_1,f_2 : A \mapsto B$ and $A_0\subseteq A$, $\mapr{f_1}{f_2}{A_0}$ iff $\forall a\in A_0, f_1(a)=f_2(a)$.
For any mappings  $f_1 : A_1 \mapsto B$ and  $f_2 : A_2 \mapsto B$,  we write $f_1[f_2]$ the operation which updates $f_1$ according to $f_2$, namely 
\[f_1[f_2](x) \deff \left\{ 
\begin{array}{l l}
  f_2(x) & \quad \text{if $x\in A_1\cap A_2$}\\
  f_1(x) & \quad \text{if $x\in A_1\setminus A_2$}\\ \end{array} \right. \]
Given $\Gamma : Var \mapsto \cL$ ,  $X\subseteq Var$ and $t\in \cL$, we write $=_{\Gamma,X,t}$ for the reflexive and symmetric relation which relates states that are equal on all variables having type $v\sqsubseteq t$ in environment $\Gamma$, provided that they are equal on all variables in $X$:  $\sigma=_{\Gamma,X,t}\sigma'$ iff $\sigma=_X\sigma' \Rightarrow \big( \forall x, ( \Gamma(x) \sbs t  \Rightarrow \sigma(x)= \sigma'(x))\big)$. When $X = \emptyset$, we omit it, hence  we write  $=_{\Gamma,t}$ instead of  $=_{\Gamma,\emptyset, t}$.
\begin{defiC}[\cite{hunt2006}]
Let $\cR$ and $\cS$ be reflexive and symmetric relations on states. We say that program $c$ maps $\cR$ into $\cS$, written $c: \cR \Longrightarrow \cS$, iff $\forall \sigma, \rho$, if $\exec{c}{\sigma}{\sigma'}$ and $\exec{c}{\rho}{\rho'}$  then $\sigma \cR \rho \Rightarrow \sigma' \cS \rho'$.
\end{defiC}

The type system we propose enjoys the following useful property:  \\
\[
\mbox{
\mbox{if}~\nntype{}{\trH{\Gamma}{c}{\Gamma'}}\;\mbox{then}\; 
     c~:~=$_{\Gamma,\Gamma(X_I)} \;\Longrightarrow\; =_{\Gamma',X_O,\alpha(X_O) \sqcup \Gamma(X_I)}$
}
\]
This property is an immediate consequence of Theorem~\ref{l:sec}. 

Hence, in order to prove that the above program $c$ is output sensitive non-interferent according to Definition~\ref{iol-sec}, it is enough to check that for all $x_l \in X_L$, $\Gamma'(x_l) \sbs \alpha(X_O) \sqcup \Gamma(X_I)$. Two executions of the program $c$ starting from  initial states that coincide on input variables $X_I$, and ending in final states that coincide on output variables $X_O$, will coincide also on the leaking variables $X_L$.

We now formally introduce our typing system.
Due to assignments, values and  types of variables  change dynamically. For example let us assume that  at some point during the execution, the value of $x$  depends on the initial value of some variable $y$ and the current value of some output variable $o$ (which itself depends on the initial value of some variable $h$), formally captured by an environment $\Gamma$ where  $\Gamma(o)=\Gamma_0(h)$ and  $\Gamma(x) = \Gamma_0(y) \js \alpha(o)$, where $\Gamma_0$ represents the initial environment. If the next to be executed instruction  is some assignment to $o$, then the current value of $o$ will change, so we have to mirror this in the new type of $x$: even if the value of $x$ does not change, its new type will be $\Gamma'(x) = \Gamma_0(y) \js \Gamma_0(h)$  (assuming that $\alpha(o)\not\sbs \Gamma_0(y)$). 
Hence $\Gamma'(x)$ is obtained by replacing in  $\Gamma(x)$ the symbolic dependency $\alpha(o)$ with the real dependency $\Gamma(o)$.
The following definition formalizes this operation that allows to replace an atom $t^0$ by another type $t'$ in a type $t$ (seen as 
the join of the atoms of its decomposition).

\begin{defi}\label{def_r}
\label{triangle}
If $t^0\in \cT_O$ is an atom and  $t',t\in \cL$ are arbitrary types, then we denote by $t[t'/t^0]$ the type obtained by replacing (if any) the occurrence of $t^0$ by $t'$ in the decomposition $At(t)$ in atoms of $t$ 
\[\displaystyle t[t'/t^0] \deff \left\{ 
\begin{array}{l l}
   t &\quad \text{if $t^0\not\in \ At(t)$}\\   
   t' \js \js_{b\in At(t)\setminus \{t^0\}}b &\quad \text{if $t^0\in \ At(t)$}\\
\end{array} \right. \]
Now we extend this definition to environments: let $x\in X_O$ and $p\in \cL$.  Then $\Gamma_1\deff \Gamma\lhd_\alpha x$ represents the environment where the symbolic dependency on the last value of $x$ of all variables  is replaced by the real type of $x$: $\Gamma_1(y) \deff (\Gamma(y))[\Gamma(x)/\alpha(x)]$.
Similarly, $(p, \Gamma) \lhd_\alpha x\deff p[\Gamma(x)/\alpha(x)]$. 
\end{defi}

\subsection{Useful basic lemmas}

The following lemma is an immediate consequence of the Assumption \ref{as1} and Definition \ref{def_r}.
\begin{lem}\label{lrs}
Let  $x\in X_O$,   $p\in \cL$ and let us denote  $\Gamma_1\deff \Gamma\lhd_\alpha x$ and  $p_1  \deff (p, \Gamma) \lhd_\alpha x$. If  $\alpha(x)\not\sbs \Gamma(x)$, then,
\begin{enumerate}
\item\label{lrs0} For any $v\in X$, $\Gamma_1(v) = (\Gamma(v), \Gamma) \lhd_\alpha x$,
\item\label{lrs1} For all variables $y\in Var$, $\alpha(x)\not\sbs \Gamma_1(y)$.
\item\label{lrs2}   $\alpha(x)\not\sbs p_1$.
\end{enumerate}
\end{lem}

We want now to extend the above definition from a single output variable $x$ to subsets $X\subseteq X_O$.
Our typing system will ensure that each generated environment $\Gamma$ will not contain {\em circular symbolic dependencies} between output variables, i.e., 
there are no output variable $o_1,o_2 \in X_O$ such that $\alpha(o_1) \sbs \Gamma(o_2)$ and  $\alpha(o_2) \sbs \Gamma(o_1)$. 
We can associate a graph $\cG(\Gamma)=(X_O, E)$ to an environment $\Gamma$, such that $(o_1,o_2)\in E$ iff $\alpha(o_1)\sbs \Gamma(o_2)$. We say that $\Gamma$ is {\bf well formed}, denoted \ac{\Gamma},  if  $\cG(\Gamma)$ is an acyclic graph. For acyclic graphs  $\cG(\Gamma)$, we define a preorder  over $X_O$,   denoted $\sbs_\Gamma$, as the transitive closure of the relation $\{(o_1,o_2) \in X_O \times X_O \ \mid \ \alpha (o_1) \sbs \Gamma (o_2)\}$, i.e. $o_1\sbs_\Gamma o_2$ iff there is a path from $o_1$ to $o_2$ in $\cG(\Gamma)$. We also define the {\it reachable variables } of $x\in X_O$ w.r.t. $\Gamma$, denoted {\bf $\gt{\Gamma}{x}$},  to be the set of all $o\in X_O$ such that $x\sbs_\Gamma o$.
	Now, for acyclic graphs  $\cG(\Gamma)$, we can extend Definition~\ref{triangle} to  subsets $X\subseteq X_O$, by first fixing an ordering $X=\{x_1,x_2,\ldots x_n\}$ of variables in $X_O$ compatible with the graph (i.e. $j\leq k$ implies that  $x_j   \not\sbs_\Gamma x_k$), and then  $(p, \Gamma) \lhd_\alpha X \deff (((p, \Gamma)\lhd_\alpha x_1 ) \lhd_\alpha x_2) \ldots \lhd_\alpha x_n$.  We also denote $\Gamma \lhd_\alpha X \deff ((\Gamma\lhd_\alpha x_1 ) \lhd_\alpha x_2) \ldots \lhd_\alpha x_n$ (in this case the ordering is not important, i.e. $(\Gamma\lhd_\alpha x_1 ) \lhd_\alpha x_2 = (\Gamma\lhd_\alpha x_2 ) \lhd_\alpha x_1$). 

The following lemma can be proved by induction on the size of $X$ using Lemma \ref{lrs}.
\begin{lem}\label{lrsA}
Let  $X\subseteq X_O$,   $p\in \cL$ and let us denote  $\Gamma_2\deff \Gamma\lhd_\alpha X$ and  $p_2  \deff (p, \Gamma) \lhd_\alpha X$. If  $\wf{\Gamma}$, then,
\begin{enumerate}
\item\label{lrsA0}  For any $v\in X$, $\Gamma_2(v) = (\Gamma(v), \Gamma) \lhd_\alpha X$,
\item\label{lrsA1} For all variables $x\in X$, and all variables $y\in Var$, $\alpha(x)\not\sbs \Gamma_2(y)$.
\item\label{lrsA2}  For all variables $x\in X$,  $\alpha(x)\not\sbs p_2$.
\end{enumerate}
\end{lem}

Next Lemma gives a precise characterization of the new preorder induced by the application of the operator $ \lhd_\alpha$.

\begin{lem}\label{lrsB}
Let  $\Gamma$  be a well formed environment and let  $x\in X_O$ and $X\subseteq X_O$. Let us denote  $\Gamma_1\deff \Gamma\lhd_\alpha x$ and   $\Gamma_2\deff \Gamma\lhd_\alpha X$. Then  $\Gamma_1$ and    $\Gamma_2$ are  well formed. Moreover, $\sbs_{\Gamma_1} =\sbs_\Gamma \setminus \{(x, o)  \mid \  o\in X_O \}$ and $\sbs_{\Gamma_2} =\sbs_\Gamma \setminus \{(x, o)  \mid \ x\in X, o\in X_O\}$.
\end{lem}

\begin{proof}
The key remark is that any edge of $\cG(\Gamma_1)$ where   $\Gamma_1\deff \Gamma\lhd_\alpha x$ corresponds to either an edge or to a path of length two in $\cG(\Gamma)$. Indeed, let $x_1,x_2\in X_O$ such that there exists an edge from $x_1$ to $x_2$ in $\cG(\Gamma_1)$, that is  $\alpha(x_1)\sbs \Gamma_1(x_2)=\Gamma(x_2)[\Gamma(x)/\alpha(x)]$.  Then either $\alpha(x_1)\sbs \Gamma(x_2)$ or $\alpha(x)\sbs \Gamma(x_2)$ and $\alpha(x_1)\sbs \Gamma(x)$. Hence either there is an edge from $x_1$ to $x_2$ in $\cG(\Gamma)$ or there must exist edges from $x_1$ to $x$ and from $x$ to $x_2$ (and hence a path of length two from $x_1$ to $x_2$)  in $\cG(\Gamma)$. 
Now the assertion of the Lemma is an immediate consequence of the above remark and Lemma \ref{lrs}. 
\end{proof}

Let $\aff(c)$ be the set of assigned variables in a program $c$, 
formally defined by:
\[\aff(c) \deff \left\{ 
\begin{array}{l l}
   \{x\} &\quad \text{if $c\equiv x:=e$}\\
   \emptyset& \quad \text{if $c\equiv skip$}\\
   \aff(c_1) \cup \aff(c_2) & \quad \text{if $c\equiv c_1;c_2$}\\
   \aff(c_1) \cup \aff(c_2) & \quad \text{if $c\equiv \iifthenelse{e}{c_1}{c_2}$}\\
   \aff(c) &\quad \text{if $c\equiv \whilep{e}{c}$}\\
\end{array} \right. \]
and let us denote $\affi(c) \deff \aff(c)\cap(Var\setminus X_O)$ and $\affo(c) \deff \aff(c)\cap X_O$.
\begin{figure*}[!tbp]
\begin{scriptsize}
	\begin{tabular}{p{0.5\linewidth} p{0.5\linewidth}}
\begin{prooftree}
\AxiomC{$x\not\in X_O$} 
   \LeftLabel{As1}
\UnaryInfC{\nntype{p}{\trH{\Gamma}{x:=e}{\Gamma[x\mapsto p\js \Gamma[\alpha](fv(e))]}}}
\end{prooftree}
&
\begin{prooftree}
\AxiomC{$x\in X_O\setminus fv(e)$}
\AxiomC{$\Gamma_1=\Gamma\lhd_\alpha x$}
   \LeftLabel{As2}
\BinaryInfC{\nntype{p}{\trH{\Gamma}{x:=e}{\Gamma_1[x \mapsto p\js \Gamma_1[\alpha](fv(e))]}}}
\end{prooftree}
\\
	\end{tabular}
	\begin{tabular}{p{0.3\linewidth} p{0.7\linewidth}}
\begin{prooftree}
 \AxiomC{}
   \LeftLabel{Skip}
\UnaryInfC{\nntype{p}{\trH{\Gamma}{skip}{\Gamma}}}
\end{prooftree}
&
\begin{prooftree}
\AxiomC{$x\in  X_O\cap fv(e)$}
\AxiomC{$\Gamma_1=\Gamma\lhd_\alpha x$}
   \LeftLabel{As3}
\BinaryInfC{\nntype{p}{\trH{\Gamma}{x:=e}{\Gamma_1[x \mapsto  p\js \Gamma(x) \sqcup \Gamma_1[\alpha](fv(e)\setminus x)]}}}
\end{prooftree}
	\end{tabular}
	\\
\begin{tabular}{p{0.34\linewidth} p{0.66\linewidth}}
\begin{prooftree}
	\AxiomC{\nntype{p}{\trH{\Gamma}{c_1}{\Gamma_1}}\hspace*{-2em}}
\AxiomC{\nntype{p}{\trH{\Gamma_1}{c_2}{\Gamma_2}}}
   \LeftLabel{Seq}
\BinaryInfC{\nntype{p}{\trH{\Gamma}{c_1;c_2}{\Gamma_2}}}
\end{prooftree}
&
\begin{prooftree}
	\AxiomC{$p_0 \sqsubseteq_r p_1$\hspace*{-2em}}
 \AxiomC{$\Gamma \sqsubseteq_r \Gamma'$\hspace*{-2em}}
 \AxiomC{\nntype{p_1}{\trH{\Gamma'}{c}{\Gamma_1'}}\hspace*{-2em}}
\AxiomC{$\Gamma_1'\sqsubseteq_r \Gamma_1$}
 \LeftLabel{Sub}
\QuaternaryInfC{\nntype{p_0}{\trH{\Gamma}{c}{\Gamma_1}}}
\end{prooftree}
 \\
\end{tabular}
\\
 \begin{tabular}{c}
\begin{minipage}[h]{\linewidth}
\begin{prooftree}
\AxiomC{\nntype{p\js p'}{\trH{\Gamma}{c_i}{\Gamma_i}}}
\AxiomC{$p'=(\Gamma[\alpha](fv(e)), \Gamma) \lhd_\alpha(\affo(c_1)\cup\affo(c_2))$}
\noLine
\UnaryInfC{$\Gamma'=  \Gamma_1 \lhd_\alpha \affo(c_2)\js \Gamma_2 \lhd_\alpha \affo(c_1)$}
  \LeftLabel{If}
\BinaryInfC{\nntype{p}{\trH{\Gamma}{\iifthenelse{e}{c_1}{c_2}}{\Gamma'}}}
\end{prooftree}
\end{minipage}
\end{tabular}
\\

%

\begin{tabular}{c}
\\

\begin{minipage}[h]{\linewidth}
\begin{prooftree}
\AxiomC{\nntype{p\js p_e}{\trH{\Gamma_1}{c}{\Gamma'}}}
\AxiomC{$p_e=\Gamma_1[\alpha](fv(e))$}
\noLine
\UnaryInfC{$(\Gamma \lhd_\alpha \affo(c))\js (\Gamma' \lhd_\alpha \affo(c)) \sbs_r \Gamma_1$}
 \LeftLabel{Wh}
\BinaryInfC{\nntype{p}{\trH{\Gamma}{\whilep{e}{c}}{\Gamma_1}}}
\end{prooftree}
\end{minipage}
\end{tabular}
\end{scriptsize}
\caption{Flow-sensitive typing rules for commands with output}\label{fig:T1}
\end{figure*}

We define the ordering over environments as usual:  $$\displaystyle\Gamma_1 \sqsubseteq \Gamma_2 \deff  \bigwedge_{x\in Var}  \Gamma_1(x) \sbs \Gamma_2(x).$$ 

We also define a restricted  ordering over environments:  $$\displaystyle\Gamma_1 \sqsubseteq_r \Gamma_2 \deff  \bigwedge_{x\in Var}  \Gamma_1(x) \sbs \Gamma_2(x) \wedge  \bigwedge_{o\in X_O, x\in Var}  (\alpha(o) \sbs \Gamma_2(x)  \Rightarrow \alpha(o) \sbs \Gamma_1(x)) .$$  

It is immediate that $\Gamma_1 \sqsubseteq_r \Gamma_2$ implies $\Gamma_1 \sqsubseteq \Gamma_2$.
Intuitively, when enriching an environment using $\sbs_r$, we have the right to add only ``real dependencies'' (and not ``symbolic'' dependencies with respect to variables in $X_O$).  We adapt this definition for elements  $t_1,t_2\in \cL$ as well: we denote $t_1\sbs_r t_2$ when  $t_1\sbs t_2$, and for all $o\in X_O$, $\alpha(o)\sbs t_2 \Rightarrow \alpha(o)\sbs t_1$, i.e. $t_2$ does not contain new symbolic dependencies w.r.t. $t_1$.

Next lemma is immediate from the definitions. 
\begin{lem}\label{pl2}
Let  $\Gamma_1, \Gamma_2, \Gamma_3$ such that $\Gamma_1\sqsubseteq \Gamma_2$, $p_1,p_2\in \cL$, $o\in X_O$ and $X\subseteq X_O$ with $p_1\sbs_r p_2$. Then $\Gamma_1 \lhd_\alpha o  \sqsubseteq \Gamma_2 \lhd_\alpha o$,  $\Gamma_1 \lhd_\alpha X  \sqsubseteq \Gamma_2 \lhd_\alpha X$, $(p_1, \Gamma_1) \lhd_\alpha o  \sqsubseteq (p_2, \Gamma_2) \lhd_\alpha o$ and   $(p_1, \Gamma_1) \lhd_\alpha X  \sqsubseteq (p_2, \Gamma_2) \lhd_\alpha X$. Moreover, $\Gamma_1 \lhd_\alpha X  \sbs_r \Gamma_3$ implies that $\Gamma_3 \lhd_\alpha X  = \Gamma_3$.
\end{lem}
The last assertion of the lemma is a consequence of the remark that $\Gamma_1 \lhd_\alpha X  \sbs_r \Gamma_3$ implies that $\Gamma_3$ does not contain more ``symbolic dependencies'' than $\Gamma_1 \lhd_\alpha X$, and  $\Gamma_1 \lhd_\alpha X$ does not contain any  ``symbolic dependencies'' with respect to variables in $X$.  Obviously, using that $\sbs_r \subseteq \sbs$, all inequalities hold also when the premise $\Gamma_1\sqsubseteq \Gamma_2$ is replaced by  $\Gamma_1\sqsubseteq_r \Gamma_2$.

\subsection{Typing rules}
\label{sec-typing-rules}

For a command $c$, judgements have the form
\nntype{p}{\trH{\Gamma}{c}{\Gamma'}} where $p\in \cL$ and $\Gamma$ and
$\Gamma'$ are type environments well-formed.  The inference rules are
shown in Figure \ref{fig:T1}. The idea is that if $\Gamma$ describes
the security levels of variables which hold before execution of $c$,
then $\Gamma'$ will describe the security levels of those variables
after execution of $c$. The type $p$ represents the usual program
counter level and serves to eliminate indirect information flows; the
derivation rules ensure that all variables that can be changed by $c$
will end up (in $\Gamma'$) with types greater than or equal to $p$. As
usual, whenever $p=\bot$ we drop it and write
\nntype{}{\trH{\Gamma}{c}{\Gamma'}} instead of
\nntype{\bot}{\trH{\Gamma}{c}{\Gamma'}}.  Throughout this paper the
type of an expression $e$ is defined simply by taking the lub of the
types of its free variables $\Gamma[\alpha](fv(e))$, for example the
type of $x+y+o$ where $o$ is the only output variable is
$\Gamma(x)\js\Gamma(y)\js\alpha(o)$. This is consistent with the
typing rules used in many systems, though more sophisticated typing rules
for expressions would be possible in principle.

Let us explain some of the typing rules:
\begin{itemize}
	\item The rule $As1$ is a standard rule for assignment, the new value of $x$ depends on the variables occuring in the right-hand side $e$.   
Since $x$ is not an output variable, we do not need to update the type of the other variables. Moreover, notice that
considering the type of an expression to be $\Gamma[\alpha](fv(e))$ instead of $\Gamma(fv(e))$ allows to capture the dependencies with
respect to the current values of output variables. 
	\item The rule $As2$ captures the fact that when assigning an output variable $x$ we need to update the types of all the other variables
		depending on the last previous value of $x$: $\Gamma_1=\Gamma\lhd_\alpha x$ express that {\em symbolic} dependencies with respect to the last previous value of $x$ 
		should be replaced with {\em real} dependencies with respect to the initial types of variables.
	\item The rule $As3$ is similar to $As2$ when the assigned variable $x$ occurs also on the right-hand side. 
	\item The rule $Sub$ is the standard subtyping rule. Notice that we use relation $\sbs_r$ instead of $\sbs$ in order to prevent introducing circular dependencies.
	\item The rule $If$ deals with conditional statements.  In an $\textsf{If }$ statement, the  program counter level changes for the typing of each branch in order to take into account the indirect 
		   information flow. Moreover, at the end of the  $\textsf{If }$ command, we do the join of the two environments obtained after the both branches,  but in order to prevent cycles, 
		    we first replace   the ``symbolic'' dependencies  by  the corresponding  ``real'' dependencies for each output variable that is assigned by the other branch.
\end{itemize}

In order to give some intuition about the rules, we present a simple example in Figure \ref{fig:ex1}. 
\begin{exa}\label{ex2:5}
Let $\{x,y,z,u\}\subseteq Var\setminus X_O$ and $\{o_1,o_2,o_3\}\subseteq X_O$ be some variables and let $\{X, Y, Z, U, O_1, O_2, O_3, \ba{O_1}, \ba{O_2}, \ba{O_3}\}$ be some atoms of a lattice of security levels. Let us assume that $\forall i\in \{1,2,3\}$, $\alpha(o_i)=\ba{O_i}$. We assume that the initial  environment is $\Gamma_0= [x\rightarrow X, y\rightarrow Y, z\rightarrow Z, u\rightarrow U, o_1\rightarrow O_1, o_2\rightarrow O_2, o_3\rightarrow O_3]$. Since the types of variables $x,u$ and $o_3$ do not change, we omit them in the following. We highlighted the changes with respect to the previous environment. 
\begin{figure*}[!tbp]
\[
\begin{array}{llr}
 & & \! \! \! \! \! \! p=\bot, \ \ \ \Gamma_0= [y\rightarrow Y, z\rightarrow Z, o_1\rightarrow O_1, o_2\rightarrow O_2]\\
(1) & o_1:=x+1 & \\
& & \! \! \! \Gamma_1= [y\rightarrow Y, z\rightarrow Z, {\bf o_1\rightarrow X}, o_2\rightarrow O_2]\\
(2) & y:=o_1+z & \\
& &  \! \! \! \Gamma_2= [{\bf  y\rightarrow \ba{O_1}\js Z}, z\rightarrow Z,  o_1\rightarrow X, o_2\rightarrow O_2]\\
(3) & o_1:=u & \\
& &  \! \! \! \Gamma_3= [{\bf  y\rightarrow X\js Z}, z\rightarrow Z,  {\bf o_1\rightarrow U}, o_2\rightarrow O_2]\\
(4) & z:=o_1+o_3 & \\
& &   \! \! \! \! \! \! \Gamma_4= [y\rightarrow X\js Z,{\bf  z\rightarrow \ba{O_1}\js\ba{O_3}},  o_1\rightarrow U, o_2\rightarrow O_2]\\
(5) &\textsf{If } (o_2=o_3+x) & {\bf  p=\ba{O_3}\js O_2\js X}\ \ \ \ \ \ \ \ \ \ \ \ \ \ \ \ \ \ \ \ \\
(6) &  \ \textsf{then} \ \ o_1:=o_2 & \\
& &  \! \! \! \! \! \!  \! \! \! \! \! \! \! \! \! \! \! \! \! \! \! \! \! \! \!  \! \! \! \! \! \! \! \Gamma_6= [y\rightarrow X\js Z,{\bf  z\rightarrow U\js\ba{O_3}}, {\bf  o_1\rightarrow \ba{O_3}\js O_2\js X\js \ba{O_2}}, o_2\rightarrow O_2]\\
(7) &  \ \textsf{else} \ \ o_2:=o_1 & \\
& &  \! \! \! \! \! \!  \! \! \!  \! \! \! \! \! \! \! \! \! \! \! \! \!  \! \! \! \! \! \! \!  \! \! \! \Gamma_7= [y\rightarrow X\js Z, z\rightarrow \ba{O_1}\js\ba{O_3}, o_1\rightarrow U, o_2\rightarrow {\bf \ba{O_3}\js O_2\js X\js \ba{O_1}}]\\
(8) &  \ \textsf{fi}  & \\
& &  \! \! \! \! \! \! \! \! \! \! \! \! \! \! \! \! \! \! \!  \! \! \! \! \! \! \! \Gamma_8= (\Gamma_6 \lhd_\alpha o_2) \js (\Gamma_7 \lhd_\alpha o_1) = [y\rightarrow X\js Z, {\bf z\rightarrow U\js\ba{O_3}},  \ \ \ \ \ \ \ \ \ \ \ \ \ \ \ \ \ \\
& & o_1\rightarrow  {\bf \ba{O_3}\js O_2\js X\js U}, o_2\rightarrow {\bf \ba{O_3}\js O_2\js X\js U}]\\
\end{array}
\]
\caption{Example of application for our typing system}\label{fig:ex1}
\end{figure*}
After the first assignment, the type of $o_1$ becomes $X$, meaning that the current value of $o_1$ depends on the initial value of $x$. After the assignment $y:=o_1+z$, the type of $y$ becomes $\ba{O_1}\js Z$, meaning that the current value of $y$ depends on the initial value of $z$ and the current value of $o_1$.  After the assignment $o_1=u$, the type of $y$ becomes $X\js Z$ as $o_1$ changed and we have to mirror this in the dependencies of $y$, and  the type of  $o_1$ becomes $X$. When we enter in the $\textsf{If }$, the  program counter level changes to $p=\ba{O_3}\js O_2\js X$ as the expression $o_2=o_3+x$ depends on the values of variables $o_2,o_3,x$, but $o_2$ and $o_3$ are output variables and $o_2$ will be assigned by the  $\textsf{If }$ command, hence we replace the ``symbolic'' dependency $\alpha(o_2)=\ba{O_2}$ by its ``real'' dependency $\Gamma(o_2)=O_2$. At the end of the  $\textsf{If }$ command, we do the join of the two environments obtained after the both branches,  but in order to prevent cycles, we first replace   the ``symbolic'' dependencies  by  the corresponding  ``real'' dependencies for each output variable that is assigned by the other branch.
\end{exa}

\subsection{Well-formed environments}

In this section we prove that, if the initial environment is well-formed, then all the environments generated by the typing system are well-formed too.
To do so, the following lemma states some  useful properties, where for any $p\in \cL$, we denote by $\mina(p)\deff \{o\in X_O \ \mid \ \alpha(o)\sbs p\}$.
\begin{lem}\label{pl1}
For all $\Gamma$, if  \wf{\Gamma} and \ \nntype{p}{\trH{\Gamma}{c}{\Gamma'}} and $\mina(p)\cap \affo(c) = \emptyset$   
then 
\begin{enumerate}
\item\label{pl1:0} \wf{\Gamma'},
\item\label{pl1:1}  for any $o\in X_O \setminus(\affo(c)\cup \mina(p))$,  $\gt{\Gamma'}{o}\subseteq \gt{\Gamma}{o}$,
\item\label{pl1:1B}  for any $o\in \mina(p)$,  $\gt{\Gamma'}{o}\subseteq \gt{\Gamma}{o}\cup \affo(c)$,
\item\label{pl1:2} for any $o\in \affo(c)$, $\gt{\Gamma'}{o}\subseteq \affo(c)$, 
\item\label{pl1:3} for any $x\not\in \aff(c)$, $(\Gamma\lhd_\alpha \affo(c))(x)  \sbs_r \Gamma'(x)$.
\end{enumerate}
\end{lem}
\begin{proof}
Proof is by induction on the derivation of \nntype{p}{\trH{\Gamma}{c}{\Gamma'}} for all assertions in the same time. 
We do a case analysis according to the last rule applied (case Skip is trivial).
\begin{itemize}[align=left,itemindent=-4mm]
\item[(Ass1)] $c$ is an assignment $x:=e$ for some $x\not\in X_O$.\\ 
Since  $x\not\in X_O$, it follows that $\cG(\Gamma')= \cG(\Gamma)$  and obviously for any  $x\not\in \aff(c)$, $\Gamma(x)=\Gamma'(x)$ and  $\sbs_{\Gamma} = \sbs_{\Gamma'}$ and  \wf{\Gamma} implies \wf{\Gamma'}.

\item[(Ass2)] $c$ is an assignment $o:=e$ for some $o\in X_O\setminus fv(e)$. Hence in this case  $\aff(c)=\affo(c)=\{o\}$, and by assumption, $\alpha(o)\not\sbs p$.

Then $\Gamma'   =\Gamma_1[o \mapsto p\js \Gamma_1[\alpha](fv(e))]$ with $\Gamma_1=\Gamma\lhd_\alpha o$.
By Lemma \ref{lrsB}, $\wf{\Gamma_1}$, and using Lemma \ref{lrs} we get that $\wf{\Gamma_1}$ does not contain any edge with origin $o$ and hence  $\wf{\Gamma'}$. The second part follows from the remark that using Lemma \ref{lrsB} we get $\sbs_{\Gamma_1} =\sbs_\Gamma \setminus \{(o, o')  \mid \  o'\in X_O \}$ and $\sbs_{\Gamma'} =\sbs_{\Gamma_1} \cup \{(o', o)  \mid \  o'\in \mina(p) \}$ , and for any $o'\not=o$, $\Gamma_1(o')=\Gamma(o')$. Hence, $\gt{\Gamma'}{o}=\emptyset$. Finally, for any  $x\not\in \aff(c)$, $(\Gamma\lhd_\alpha o)(x) =  \Gamma_1 (x)=\Gamma'(x)$.

\item[(Ass3)]  $c$ is an assignment $o:=e$ for some $o\in X_O\cap fv(e)$.  Similar to the previous case.
\item[(Seq)]  
$c$ is $c_1;c_2$.
Then \nntype{p}{\trH{\Gamma}{c_1;c_2}{\Gamma_2}} was inferred based on 
\nntype{p}{\trH{\Gamma}{c_1}{\Gamma_1}}
and \nntype{p}{\trH{\Gamma_1}{c_2}{\Gamma_2}}. Let us denote $U_i=\affo(c_i)$, for  $i=1,2$.

 Using  the induction  hypothesis for \nntype{p}{\trH{\Gamma}{c_1}{\Gamma_1}} we get that 
 $\wf{\Gamma_1}$, and for any $o\in X_O\setminus( U_1\cup \mina(p))$, $\gt{\Gamma_1}{o}\subseteq \gt{\Gamma}{o}$,  for any $o\in \mina(p)$, $\gt{\Gamma_1}{o}\subseteq \gt{\Gamma}{o}\cup  U_1$  and for any $o\in U_1$, $\gt{\Gamma_1}{o}\subseteq U_1$. In addition,  for any $x\not\in  U_1$, $(\Gamma\lhd_\alpha U_1))(x)  \sbs_r \Gamma_1(x)$.
 
 Using  the induction  hypothesis for \nntype{p}{\trH{\Gamma_1}{c_2}{\Gamma_2}} we get that 
 $\wf{\Gamma_2}$, and for any $o\in X_O\setminus (U_2\cup \mina(p))$, $\gt{\Gamma_2}{o}\subseteq \gt{\Gamma_1}{o}$,  for any $o\in \mina(p)$, $\gt{\Gamma_2}{o}\subseteq \gt{\Gamma_1}{o}\cup  U_2$ and for any $o\in U_2$, $\gt{\Gamma_2}{o}\subseteq U_2$. In addition,  for any $x\not\in  U_2$, $(\Gamma_1 \lhd_\alpha U_2)(x)  \sbs_r \Gamma_2(x)$.
 
 Since $ \affo(c)=U_1\cup U_2$,  for any  $o\in X_O\setminus (U_1\cup U_2\cup \mina(p))$, we have both  $o\in X_O\setminus (U_1\cup \mina(p))$ and $o\in X_O\setminus (U_2\cup \mina(p))$ and hence  $\gt{\Gamma_2}{o}\subseteq \gt{\Gamma_1}{o}\subseteq \gt{\Gamma}{o}$. 

For any  $o\in \mina(p)$, $\gt{\Gamma_2}{o}\subseteq \gt{\Gamma_1}{o}\cup  U_2 \subseteq  \gt{\Gamma}{o}\cup  U_1 \cup  U_2$.

Now, for any  $o\in  (U_1\cup U_2)$,  either $o\in U_2$ or  $o\in U_1\setminus U_2$. If $o\in U_2$, then  $\gt{\Gamma_2}{o}\subseteq U_2  \subseteq U_1\cup U_2$. If $o\in U_1\setminus U_2$, then  $\gt{\Gamma_2}{o}\subseteq  \gt{\Gamma_1}{o} \subseteq U_1 \subseteq U_1 \cup U_2$.

Finally, for any  $x\not\in \aff(c)$,  we have   $(\Gamma\lhd_\alpha (U_1 \cup  U_2)(x)  = ((\Gamma\lhd_\alpha U_1)\lhd_\alpha U_2)(x) = ((\Gamma\lhd_\alpha U_1)(x), \Gamma\lhd_\alpha U_1)  \lhd_\alpha U_2    \stackrel{(1)}\sbs_r  ((\Gamma\lhd_\alpha U_1)(x), \Gamma_1)  \lhd_\alpha U_2    \stackrel{(2)}\sbs_r  (\Gamma_1(x), \Gamma_1) \lhd_\alpha U_2 = (\Gamma_1 \lhd_\alpha U_2)(x)   \stackrel{(3)}\sbs_r  \Gamma_2(x)$. We used whenever necessarily Lemma \ref{pl1}; in addition, in (1) we used that $(\Gamma\lhd_\alpha U_1)(x)$ does not depend on variables in $U_1$, and by induction hypothesis for all variables $v\not\in  U_1$, $(\Gamma\lhd_\alpha U_1)(v)  \sbs_r \Gamma_1(v)$, in (2) we used that  by induction hypothesis  $(\Gamma\lhd_\alpha U_1)(x)  \sbs_r \Gamma_1(x)$, in (3) we used that  $x\not\in  U_2$, and hence by induction hypothesis, $(\Gamma_1 \lhd_\alpha U_2)(x)  \sbs_r \Gamma_2(x)$.
\item[(If)] 
$c$ is $\iifthenelse{e}{c_1}{c_2}$ . 
Then \nntype{p}{\trH{\Gamma}{\iifthenelse{e}{c_1}{c_2}}{\Gamma'}} was inferred based on 
\nntype{p\js p'}{\trH{\Gamma}{c_i}{\Gamma_i}}, where 
$p'=(\Gamma[\alpha](fv(e)), \Gamma) \lhd_\alpha(\affo(c_1)\cup\affo(c_2))$
and 
$\Gamma'= \Gamma'_1 \js \Gamma'_2$ where $\Gamma'_1 =  \Gamma_1 \lhd_\alpha \affo(c_2)$ and $\Gamma'_2= \Gamma_2 \lhd_\alpha \affo(c_1)$. Let us denote $U_i=\affo(c_i)$, for  $i=1,2$. First notice that  $\mina(p)\cap (U_1\cup U_2) = \emptyset$ ensures that  $\mina(p\js p')\cap (U_1\cup U_2) = \emptyset$ .

 Using  the induction  hypothesis for \nntype{p\js p'}{\trH{\Gamma}{c_1}{\Gamma_1}} we get that 
 $\wf{\Gamma_1}$, and for any $o\in X_O\setminus( U_1\cup \mina(p\js p'))$, $\gt{\Gamma_1}{o}\subseteq \gt{\Gamma}{o}$,  for any $o\in \mina(p\js p')$, $\gt{\Gamma_1}{o}\subseteq \gt{\Gamma}{o}\cup  U_1$  and for any $o\in U_1$, $\gt{\Gamma_1}{o}\subseteq U_1$. In addition,  for any $x\not\in  U_1$, $(\Gamma\lhd_\alpha U_1))(x)  \sbs_r \Gamma_1(x)$.

 Using  the induction  hypothesis for \nntype{p\js p'}{\trH{\Gamma}{c_2}{\Gamma_2}} we get that 
 $\wf{\Gamma_2}$, and for any $o\in X_O\setminus( U_2\cup \mina(p\js p')$, $\gt{\Gamma_2}{o}\subseteq \gt{\Gamma}{o}$,  for any $o\in \mina(p\js p')$, $\gt{\Gamma_2}{o}\subseteq \gt{\Gamma}{o}\cup  U_2$  and for any $o\in U_2$, $\gt{\Gamma_2}{o}\subseteq U_2$. In addition,  for any $x\not\in  U_2$, $(\Gamma\lhd_\alpha U_2))(x)  \sbs_r \Gamma_2(x)$.

For any $o\in X_O\setminus (U_1\cup U_2\cup \mina(p\js p'))$, $\gt{\Gamma'_1\cup \Gamma'_2}{o} =\gt{(\Gamma_1 \lhd_\alpha U_2)\cup (\Gamma_2\lhd_\alpha U_1)}{o} \subseteq \gt{\Gamma}{o}$ since both $\gt{\Gamma_1 \lhd_\alpha U_2}{o} \subseteq \gt{\Gamma_1}{o}  \subseteq  \gt{\Gamma}{o}$ and  $\gt{\Gamma_2 \lhd_\alpha U_1}{o} \subseteq \gt{\Gamma_2}{o}  \subseteq  \gt{\Gamma}{o}$.

For any  $o\in U_1$, $\gt{\Gamma_2\lhd_\alpha U_1}{o} = \emptyset$ and by induction  hypothesis  $\gt{\Gamma_1\lhd_\alpha U_2}{o} \subseteq \gt{\Gamma_1}{o}\subseteq U_1$. 
For any  $o\in U_2$, $\gt{\Gamma_1\lhd_\alpha U_2}{o} = \emptyset$ and by induction  hypothesis  $\gt{\Gamma_2\lhd_\alpha U_1}{o} \subseteq \gt{\Gamma_2}{o}\subseteq U_2$.  This implies that for any  $o\in  (U_1\cup U_2)$,  $\gt{\Gamma'_1\cup \Gamma'_2}{o} =\gt{(\Gamma_1 \lhd_\alpha U_1)\cup (\Gamma_2\lhd_\alpha U_2)}{o} \subseteq U_1 \cup U_2$.

For any  $o\in \mina(p\js p')$, by induction  hypothesis  $\gt{\Gamma_1\lhd_\alpha U_2}{o} \subseteq \gt{\Gamma_1}{o}\subseteq \gt{\Gamma}{o}\cup  U_1$ and   $\gt{\Gamma_2\lhd_\alpha U_1}{o} \subseteq \gt{\Gamma_2}{o}\subseteq \gt{\Gamma}{o}\cup U_2$.  We already proved that for any  $o\in  (U_1\cup U_2)$,  $\gt{\Gamma'_1\cup \Gamma'_2}{o} \subseteq U_1 \cup U_2$. The previous inclusions  imply that for any   $o\in \mina(p\js p')$,  $\gt{\Gamma'_1\cup \Gamma'_2}{o} \subseteq \gt{\Gamma}{o}\cup U_1 \cup U_2$.

Now if we assume by contradiction that $\neg \wf{\Gamma'}$, we get that there must exist $x_1,x_2\in X_O$ such that  $x_1\sbs_{\Gamma'_1\js \Gamma'_2} x_2$ and $x_2\sbs_{\Gamma'_1\js \Gamma'_2} x_1$. We make an analysis by case:
 \begin{itemize}
 \item $x_1\in U_1 \cap U_2$. \  Impossible, since we have that $\wf{\Gamma_1 \lhd_\alpha U_2}$ and $\wf{\Gamma_2 \lhd_\alpha U_1}$, and for any $o\in X_O$, $x_1\not\sbs_{\Gamma_1 \lhd_\alpha U_2} o$ and  $x_1\not\sbs_{\Gamma_2 \lhd_\alpha U_1} o$.


\item $x_1, x_2\in U_1 \setminus U_2$. \   From induction hypothesis we get $\gt{\Gamma'_1}{x_i}\subseteq U_1$ and $\gt{\Gamma'_2}{x_i}=\emptyset$. This implies that for any $x\in \gt{\Gamma'_1}{x_i}$, it holds $x\in U_1$ and hence $\gt{\Gamma'_2}{x}=\emptyset$.
It means that $x_1\sbs_{\Gamma'_1\js \Gamma'_2} x_2$ and $x_2\sbs_{\Gamma'_1\js \Gamma'_2} x_1$ implies that $x_1\sbs_{\Gamma'_1} x_2$ and $x_2\sbs_{\Gamma'_1} x_1$ which contradicts $\wf{\Gamma'_1}$.


\item $x_1\in U_1 \setminus U_2$ and $x_2\not\in U_1$. \   From induction hypothesis we get $\gt{\Gamma'_1}{x_1}\subseteq U_1$ and $\gt{\Gamma'_2}{x_1}=\emptyset$.  This implies that for any $x\in \gt{\Gamma'_1}{x_1}$, it holds $x\in U_1$ and hence $\gt{\Gamma'_2}{x}=\emptyset$. It means that $x_1\sbs_{\Gamma'_1\js \Gamma'_2} x$ implies that $x\in \gt{\Gamma'_1}{x_1}\subseteq U_1$, contradiction with  $x_2\not\in U_1$ and  $x_1\sbs_{\Gamma'_1\js \Gamma'_2} x_2$.


\item $x_1\not\in U_1 \cup U_2$ and  $x_2\not\in U_1 \cup U_2$. \  In this case, by induction 
$\gt{\Gamma'_1\js \Gamma'_2}{x_1}\subseteq \gt{\Gamma}{x_1}  \cup (U_1\cup U_2)$ and  $\gt{\Gamma'_1\js \Gamma'_2}{x_2}\subseteq \gt{\Gamma}{x_2} \cup (U_1\cup U_2) $. Hence $x_1\sbs_{\Gamma'_1\js \Gamma'_2} x_2$ implies that $x_1\sbs_{\Gamma} x_2$, 
$x_2\sbs_{\Gamma'_1\js \Gamma'_2} x_1$ implies that $x_2\sbs_{\Gamma} x_1$, contradiction with $\wf{\Gamma}$.

\item The remaining cases are symmetrical ones with the previous cases.
 \end{itemize}
\item[(While)] Similar to the rule (If).
\item[(Sub)] Trivial from the premises of the rule, using the induction hypothesis, the transitivity of $\sbs_r$ and that $\Gamma_1 \sbs_r \Gamma_2$  implies that   $\cG(\Gamma_1)= \cG(\Gamma_2)$.
\qedhere
\end{itemize}
\end{proof}

\subsection{Soundness of the typing system}

As already stated above, our type system aims to capture the following
non-interference condition: given a derivation
\nntype{p}{\trH{\Gamma}{c}{\Gamma'}}, the final value of a variable
$x$ with final type $t\js \alpha(X_O)$, should depend at most on the
initial values of those variables $y$ with initial types
$\Gamma(y)\sbs t$ and on the final values of variables in $X_O$. Or
otherwise said, executing a program $c$ on two initial states $\sigma$
and $\rho$ such that $\sigma(y)=\rho(y)$ for all $y$ with
$\Gamma(y)\sbs t$ which ends with two final states $\sigma'$ and
$\rho'$ such that $\sigma'(o)=\rho'(o)$ for all $o\in X_O$ will
satisfy $\sigma'(x)=\rho'(x)$ for all $x$ with $\Gamma'(x)\sbs t\js
\alpha(X_O)$.  In order to prove the soundness of the typing system,
we need a stronger invariant denoted $\cI(t,\Gamma)$: intuitively,
$(\sigma, \rho)\in \cI(t,\Gamma)$ means that for each variable $x$ and
$A\subseteq X_O$, if $\sigma =_A \rho$ and $\Gamma(x)\sbs t\js \alpha(A)$, then
$\sigma(x)=\rho(x)$. 
Formally, given $t\in \cL$ and $\Gamma : Var \mapsto \cL$, we define
$\displaystyle\cI(t,\Gamma) \deff  \bigcap_{A\subseteq X_O} =_{\Gamma, A,  \alpha(A) \sqcup t}.$

The following lemmas provide some useful properties satisfied by the invariant $\cI(t,\Gamma)$.
\begin{lem}\label{p0}
If $\Gamma_1\sqsubseteq \Gamma_2$ then for all $t\in \cL$, $\cI(t,\Gamma_1) \subseteq \cI(t,\Gamma_2)$.
\end{lem}
\begin{proof}
Assume  $(\sigma, \rho)\in\cI(t,\Gamma_1)$. We prove that  $(\sigma, \rho)\in\cI(t,\Gamma_2)$.
Let $A\subseteq X_O$ and let $y\in Var$ such that $\Gamma_2(y)\sbs  \alpha(A)  \sqcup t$.  Assume that $\sigma =_A  \rho$. We have to prove that   $\sigma(y)=\rho(y)$. We have  $\Gamma_1(y) \sbs_r \Gamma_2(y)\sbs  \alpha(A)  \sqcup t$, and since $(\sigma, \rho)\in\cI(t,\Gamma_1)$ and  $\sigma =_A  \rho$, we get  $\sigma(y)=\rho(y)$. 
\end{proof}

\begin{lem}\label{l:pwf}
Let  $x\in X_O,  X\subseteq X_O$ 
 and let  $\Gamma$ be well-formed. Let $\Gamma_1= \Gamma\lhd_\alpha x$ and  $\Gamma_2= \Gamma\lhd_\alpha X$.
 Then for all $t\in \cL$, it holds $\cI(t,\Gamma) \subseteq \cI(t,\Gamma_1)$ and $\cI(t,\Gamma) \subseteq \cI(t,\Gamma_2)$.
\end{lem}
\begin{proof}
	We prove only the first inclusion, the second one can be easily proved by induction using the first one.

Let $\sigma,  \rho$ such that  $(\sigma, \rho)\in\cI(t,\Gamma)$ and let us prove that  $(\sigma, \rho)\in\cI(t,\Gamma_1)$.

Let $B\subseteq X_O$ and let $y\in Var$ such that $\Gamma_1(y)\sbs  \alpha(B)  \sqcup t$.  Assume that $\sigma =_B  \rho$. We have to prove that   $\sigma(y)=\rho(y)$.


Then  $ \Gamma_1(y) \sbs    \alpha(B) \sqcup t$  implies that  $\big(\Gamma(y)\big)[\Gamma(x)/\alpha(x)]\sbs    \alpha(B) \sqcup t$. That is,  either $\Gamma(y) \sbs \alpha(B)\sqcup t$ or  $\Gamma(y) \sbs \alpha(B)  \sqcup \alpha(x) \sqcup t$ and $\Gamma(x)\sbs \alpha(B) \sqcup t$. 
\begin{itemize}

\item If  $\Gamma(y) \sbs \alpha(B) \sqcup t$, since  $\sigma =_{B}\rho$ and $(\sigma, \rho)\in\cI(t,\Gamma)$ we get $\sigma(y)=\rho(y)$.

\item If  $\Gamma(y) \sbs \alpha(B) \sqcup \alpha(x) \sqcup t$ and $\Gamma(x)\sbs \alpha(B) \sqcup t$, from $\sigma =_{B}\rho$ and $(\sigma, \rho)\in\cI(t,\Gamma)$, we get $\sigma =_{\{x\}}\rho$, and hence  $\sigma =_{B\cup\{x\}}\rho$. Since  $\Gamma(y) \sbs \alpha(B) \sqcup \alpha(x) \sqcup t$ and  $\sigma =_{\Gamma, B\cup\{x\}, \alpha(B\cup\{x\})  \sqcup t}\rho$, we get $\sigma(y)=\rho(y)$.
\qedhere
\end{itemize}
\end{proof}

\begin{lem}\label{pl5}
Let $o\in X_O$ and let $\Gamma' = \Gamma \lhd_\alpha o$ with well formed \wf{\Gamma}.
Let $t\in \cL$ 
 and let $(\sigma, \rho)\in \cI(t,\Gamma).  $
%
 For any  $A\subseteq X_O$ such that $\sigma =_{A\setminus\{o\}}  \rho$ and for any $y\in Var$ such that  $\Gamma'(y)\sbs \alpha(A) \sqcup t$ it holds that $\sigma(y)=\rho(y)$.
\end{lem}
\begin{proof}
\wf{\Gamma} and $ \Gamma'(y) \sbs    \alpha(A) \sqcup t$  implies that  $\big(\Gamma(y)\big)[\Gamma(o)/\alpha(o)]\sbs    \alpha(A\setminus\{o\}) \sqcup t$.
That is,  either $\Gamma(y) \sbs \alpha(A\setminus\{o\})\sqcup t$ or  $\Gamma(y) \sbs \alpha(A) \sqcup t$ and $\Gamma(o)\sbs \alpha(A\setminus\{o\}) \sqcup t$. 
\begin{itemize}
\item If  $\Gamma(y) \sbs \alpha(A\setminus\{o\}) \sqcup t$, as  $\sigma =_{A\setminus\{o\}}\rho$ and $(\sigma, \rho)\in\cI(t,\Gamma)$ we get $\sigma(y)=\rho(y)$.
\item If  $\Gamma(y) \sbs \alpha(A) \sqcup t$ and $\Gamma(o)\sbs \alpha(A\setminus\{o\}) \sqcup t$, from $\sigma =_{A\setminus\{o\}}\rho$ and $(\sigma, \rho)\in\cI(t,\Gamma)$, we get $\sigma =_{\{o\}}\rho$, and hence  $\sigma =_{A}\rho$. Since  $\Gamma(y) \sbs \alpha(A) \sqcup t$ and  $\sigma =_{\Gamma, A, \alpha(A)  \sqcup t}\rho$, we get $\sigma(y)=\rho(y)$.
\qedhere
\end{itemize}
\end{proof}
\begin{lem}\label{l:pl4}
Let $\Gamma' =  \Gamma \lhd_\alpha U$ for a well formed $\wf{\Gamma}$ where  $U=\{o_1,\ldots ,o_n\} \subseteq X_O$ .
Let $t\in \cL$  and let $(\sigma, \rho)\in \cI(t,\Gamma)$.
For any $A\subseteq X_O$ such that  $\sigma =_{A\setminus U}  \rho$ and for any $y\in Var$ such that $\Gamma'(y)\sbs \alpha(A) \sqcup t$ it holds that $\sigma(y)=\rho(y)$.
\end{lem}
\begin{proof}
By induction on $n$. The case $n=1$ follows from the lemma \ref{pl5}. Let us denote $U_{n-1}= U\setminus\{o_n\}=\{o_1,\ldots ,o_{n-1}\}$. and let $\Gamma_{n-1} =  \Gamma \lhd_\alpha U_{n-1}$. Hence $\Gamma' =  \Gamma_{n-1} \lhd_\alpha o_n$.

By lemma \ref{lrsB},  $\Gamma_{n-1}$ is  well formed too. \wf{\Gamma_{n-1}} and $ \Gamma'(y) \sbs    \alpha(A) \sqcup t$  implies that  $\big(\Gamma_{n-1}(y)\big)[\Gamma_{n-1}(o_n)/\alpha(o_n)]\sbs    \alpha(A\setminus\{o_n\}) \sqcup t$. That is,  either $\Gamma_{n-1}(y) \sbs \alpha(A\setminus\{o_n\})\sqcup t$ or  $\Gamma_{n-1}(y) \sbs \alpha(A) \sqcup t$ and $\Gamma_{n-1}(o_n)\sbs \alpha(A\setminus\{o_n\}) \sqcup t$. 
\begin{itemize}
\item If  $\Gamma_{n-1}(y) \sbs \alpha(A\setminus\{o_n\}) \sqcup t$, since  $\sigma =_{(A\setminus\{o_n\})\setminus U_{n-1}}\rho$, by induction (taking  $U'=\{o_1,\ldots ,o_{n-1}\}$ and $A'=A\setminus\{o_n\}$) we get $\sigma(y)=\rho(y)$.
\item If  $\Gamma_{n-1}(y) \sbs \alpha(A) \sqcup t$ and $\Gamma_{n-1}(o_n)\sbs \alpha(A\setminus\{o_n\}) \sqcup t$, since $\sigma =_{(A\setminus\{o_n\})\setminus U_{n-1}}\rho$, we get by induction (taking  $U'=U_{n-1}$ and $A'=A\setminus\{o_n\}$) that $\sigma =_{\{o_n\}}\rho$, and hence  $\sigma =_{A\setminus U_{n-1}}\rho$. From  $\Gamma_{n-1}(y) \sbs \alpha(A) \sqcup t$  and   $\sigma =_{A\setminus U_{n-1}}\rho$ using the induction again (taking  $U'=U_{n-1}$ and $A'=A$) we get $\sigma(y)=\rho(y)$.
\qedhere
\end{itemize}
\end{proof}

The following theorem states the soundness of our typing system.  

\begin{thm}\label{l1}
Let us assume that  \wf{\Gamma} 
and  $\forall o\in X_O, \alpha(o)\not\sbs t$. If \nntype{p}{\trH{\Gamma}{c}{\Gamma'}} then $c~:~\cI(t,\Gamma) \Longrightarrow \cI(t,\Gamma').$
\end{thm}

\noindent\begin{proof}
Proof is by induction on the derivation of \nntype{p}{\trH{\Gamma}{c}{\Gamma'}}. 
Let $\sigma, \sigma', \rho, \rho'$ such that  $(\sigma, \rho)\in\cI(t,\Gamma)$ and $\exec{c}{\sigma}{\sigma'}$ and  $\exec{c}{\rho}{\rho'}$.
We prove that $(\sigma', \rho')\in\cI(t,\Gamma')$.

Let $A\subseteq X_O$ and let $y\in Var$ such that $\Gamma'(y)\sbs  \alpha(A)  \sqcup t$.  Assume that $\sigma' =_A  \rho'$. We have to prove that   $\sigma'(y)=\rho'(y)$.


We do a case analysis according to the last rule applied (case Skip is trivial).
\begin{itemize}[align=left,itemindent=-4mm]
\item[(Ass1)] $c$ is an assignment $x:=e$ for some $x\not\in X_O$.\\ 
Then $\Gamma'   =\Gamma[x\mapsto p\js \Gamma[\alpha](fv(e))]$. Hence $\sigma' =_{A} \rho'$  implies $\sigma =_{A} \rho$.

If $y\not\equiv x$, then $\Gamma(y) = \Gamma'(y)\sbs    \alpha(A) \sqcup t$, and since  $\sigma =_{\Gamma, A, \alpha(A)  \sqcup t}\rho$, we get $\sigma'(y) = \sigma(y)=\rho(y)=\rho'(y)$. 
Let us assume that  $c$ is  $y:=e$ for some $e$ and   $y\not\in X_O$. Then $ \Gamma[\alpha](fv(e))\sbs \Gamma'(y) \sbs\alpha(A)  \sqcup t$, and this implies that:  1) for all variables $v\in fv(e)\setminus X_O$, $\Gamma(v)\sbs \alpha(A) \sqcup  t$, hence  $\sigma(v)=\rho(v)$ and 2) for all variables $v\in fv(e)\cap X_O$, $\alpha(v)\sbs \alpha(A) \sqcup  t$, hence  $v\in A$ and $\sigma(v)=\rho(v)$. We get that  $\sigma(e)=\rho(e)$, and hence  $\sigma'(y)=\rho'(y)$.
\item[(Ass2)] $c$ is an assignment $o:=e$ for some $o\in X_O\setminus fv(e)$. 

Using  Lemma \ref{l:pwf}, we get that $(\sigma, \rho)\in\cI(t,\Gamma_1)$, where $\Gamma_1=\Gamma\lhd_\alpha o$ and $\Gamma'=\Gamma_1[o \mapsto  p\js \Gamma_1[\alpha](fv(e)\setminus o)]$.

If $y\not\equiv o$, then $\Gamma_1(y) = \Gamma'(y)\sbs    \alpha(A) \sqcup t$, and since  $\sigma =_{\Gamma_1, A, \alpha(A)  \sqcup t}\rho$, we get $\sigma'(y) = \sigma(y)=\rho(y)=\rho'(y)$. 

If $y\equiv o$, then  $p\js \Gamma_1[\alpha](fv(e)\setminus o) = \Gamma'(o) \sbs \alpha(A) \sqcup  t$,    and this implies that:  1) for all variables $v\in fv(e)\setminus X_O$, $\Gamma_1(v)\sbs \alpha(A) \sqcup  t$, hence  $\sigma(v)=\rho(v)$ and 2) for all variables $v\in fv(e)\cap X_O$, $\alpha(v)\sbs \alpha(A) \sqcup  t$, hence  $v\in A$ and $\sigma(v)=\rho(v)$. We get that  $\sigma(e)=\rho(e)$, and hence  $\sigma'(y)=\rho'(y)$.
\item[(Ass3)]  $c$ is an assignment $o:=e$ for some $o\in X_O\cap fv(e)$.  Similar to the previous case, using the remark that $\Gamma(o) \sbs \alpha(A) \sqcup  t$ implies that  $\sigma(o)=\rho(o)$. 
\item[(Seq)]  Trivial, using the transitivity of $\Longrightarrow$. 
\item[(If)] Let us denote $U_i=\{o^i_1,o^i_2,\ldots ,o^i_n\}=\affo(c_i)$ for $i=1,2$  for some good orderings $U_i$ of $\affo(c_i)$. 
\begin{itemize}

\item If $(\Gamma[\alpha](fv(e)), \Gamma) \lhd_\alpha (U_1\cup U_2)\not\sbs\Gamma'(y)$, obviously we get  $y\not\in \aff(c)$, i.e.  $c$ contains no assignments to $y$. Hence  $\sigma'(y) = \sigma(y)$ and $\rho(y)=\rho'(y)$. 
Then from Lemma \ref{l:pwf}, we get that $ (\Gamma \lhd_\alpha ( U_1\cup U_2))(y) \sbs_r  \Gamma'(y) \sbs  \alpha(A) \sqcup t$.  
Moreover,  $\sigma' =_{A}\rho'$ implies that  $\sigma =_{A\setminus (U_1\cup U_2)}\rho$ and using Lemma \ref{l:pl4} we get $\sigma(y)=\rho(y)$, and hence $\sigma'(y) = \sigma(y) = \rho(y)=\rho'(y)$.

\item Let us suppose that   $(\Gamma[\alpha](fv(e)), \Gamma) \lhd_\alpha (U_1\cup U_2)\sbs\Gamma'(y)$. We denote  $\Gamma^i_{0} =  \Gamma \lhd_\alpha U_i$. 
Then $(\Gamma[\alpha](fv(e)), \Gamma) \lhd_\alpha (U_1\cup U_2) \sbs \Gamma'(y) \sbs \alpha(A) \sqcup t$, and this implies that $(\Gamma[\alpha](fv(e)), \Gamma) \lhd_\alpha (U_1\cup U_2) \sbs \alpha(A\setminus (U_1\cup U_2)) \sqcup t$.  Moreover,  $\sigma' =_{A}\rho'$ implies that  $\sigma =_{A\setminus (U_1\cup U_2)}\rho$.  Using Lemma \ref{l:pl4}, we get for each variable $v\in fv(e)$ that  $\sigma(v)=\rho(v)$. This proves that $\sigma(e)=\rho(e)$ and hence  both executions $\exec{c}{\sigma}{\sigma'}$ and  $\exec{c}{\rho}{\rho'}$ take the same branch $i$. Then we use the induction hypothesis applied to  \nntype{p\js p'}{\trH{\Gamma}{c_i}{\Gamma_i}} to get that $(\sigma', \rho')\in \cI(t,\Gamma_i)$, and then, since $\Gamma'= \Gamma_1 \lhd_\alpha U_2  \js \Gamma_2 \lhd_\alpha U_1$,  we use Lemmas \ref{p0} and \ref{l:pwf} to conclude  $(\sigma', \rho')\in \cI(t,\Gamma')$. 

\end{itemize}
\item[(While)] $c$ is $\whilep{e}{c_1}$.
  Then \nntype{p}{\trH{\Gamma}{c}{\Gamma'}} was inferred based on
  
\nntype{p\js p_e}{\trH{\Gamma'}{c}{\Gamma_1}} and 
$(\Gamma \lhd_\alpha U)\js (\Gamma_1 \lhd_\alpha U) \sbs_r \Gamma'$
where $U= \affo(c)$ and $p_e=\Gamma'[\alpha](fv(e))$.
From  $(\sigma, \rho)\in\cI(t,\Gamma)$, using Lemma \ref{l:pwf}, we get   $(\sigma, \rho)\in\cI(t,\Gamma\lhd_\alpha U)$  and using now Lemma \ref{p0} and the inequality $\Gamma\lhd_\alpha U \sbs (\Gamma \lhd_\alpha U)\js (\Gamma_1 \lhd_\alpha U) \sbs_r \Gamma'$,  we obtain  $(\sigma, \rho)\in\cI(t,\Gamma')$. Using the induction hypothesis applied to \nntype{p\js p_e}{\trH{\Gamma'}{c}{\Gamma_1}}, we get that  $(\sigma', \rho')\in\cI(t,\Gamma_1)$. We apply again  Lemma \ref{l:pwf} and we get  $(\sigma', \rho')\in\cI(t,\Gamma_1 \lhd_\alpha U)$, and using  Lemma \ref{p0} and that $(\Gamma \lhd_\alpha U)\js (\Gamma_1 \lhd_\alpha U) \sbs_r \Gamma'$, we obtain $(\sigma', \rho')\in\cI(t,\Gamma')$.

\item[(Sub)] Trivial, from Lemma \ref{p0} and the induction hypothesis.
\qedhere
\end{itemize}
\end{proof}
\subsection{Soundness w.r.t. to output-sensitive non-interference}\label{s:iol}

In this section we show how we can use the typing system in order to prove that a program $c$ is output-sensitive noninterferent.
Let $\evar = Var \cup \{\overline{o} \ | \ o\in X_O \}$.
Let us define  $\cL \deff \{\tau_A \ \mid \ A\subseteq \evar\}$. We denote $\bot = \tau_\emptyset$ and $\top = \tau_{\evar}$ and we consider the lattice $(\cL, \bot , \top, \sbs)$  with $\tau_A \js \tau_{A'} \deff \tau_{A\cup A'}$ and  $\tau_A \sbs \tau_{A'}$ iff  $A\subseteq A'$. 
Obviously, $\cL$ is a bounded atomistic lattice, its set of atoms being $\{\tau_x \ \mid \ x\in Var\} \cup \{\tau_{\overline{o}} \ | \ o\in X_O \}$.

The following Theorem is a consequence of the Definition \ref{iol-sec} and Theorem \ref{l1}.  

\begin{thm}\label{l:sec}
Let $\cL$  be the lattice described above. 
Let $(\Gamma,\alpha)$ be defined by $\Gamma(x)=\{\tau_x\}$, for all $x\in Var$ and  $\alpha(o)=\{\tau_{\overline{o}}\}$, for all $o\in X_O$.   If \ntype{p}{\trH{\Gamma}{c}{\Gamma'}}  and for all $x_l\in X_L$, $\Gamma'(x_l) \sbs \Gamma(X_I) \js \alpha(X_O)$, then $c$ is ($X_I, X_O, X_L$)-secure.
\end{thm}
\begin{proof}
Let  $t= \Gamma(X_I)$.  First, we prove that if $\sigma =_{X_I} \rho $, then $(\sigma, \rho)\in\cI(t,\Gamma)$. Let $A\subseteq X_O$ such that  $\sigma =_{A} \rho $ and let $y\in Var$ such that $\Gamma(y)\sbs  \alpha(A)  \sqcup t = \alpha(A)  \sqcup \Gamma(X_I)$.  This implies that $y\in X_I$, and since $\sigma =_{X_I} \rho$, we get $\sigma(y)=\rho(y)$.

Now let  $\sigma, \sigma', \rho, \rho'$, such $\exec{c}{\sigma}{\sigma'}$ and $\exec{c}{\rho}{\rho'}$ and  $\sigma =_{X_I} \rho $ and $\sigma'=_{X_O}\rho'$. Let $x_l\in X_L$. We have to prove that  $\sigma'=_{X_l}\rho'$. Let us apply the Theorem \ref{l1} with  $t= \Gamma(X_I)$. Since \ntype{p}{\trH{\Gamma}{c}{\Gamma'}} and  $(\sigma, \rho)\in\cI_1(t,\Gamma)$, we get that  $(\sigma', \rho')\in\cI(t,\Gamma')$. It means that $\sigma'=_{\Gamma', X_O,  \alpha(X_O) \sqcup \Gamma(X_I)}\rho'$.  Since by hypothesis we have that $\sigma'=_{X_O}\rho'$ and $\Gamma'(x_l) \sbs \alpha(X_O) \sqcup \Gamma(X_I)$, we get that $\sigma'=_{x_l}\rho'$. 
\end{proof}

\section{Output-sensitive constant-time}\label{s:3}

In this section we illustrate how our approach can be applied to a more realistic setting, considering  a specific side-channel leakage
due to the cache usage. However, this approach could be applied to any other side-channel setting as soon as one can model the leakage 
produced by each command.
 
\begin{figure*}[!tbp]
\begin{scriptsize}

\begin{tabular}{ll}
\begin{minipage}[h]{.3\linewidth}
\begin{prooftree} 
\AxiomC{$act \equiv \rd{\sigma(\vc{f})}$}
\UnaryInfC{$(x:=e,\sigma)\reda{act} \sigma[(x,0)  \mapsto\sigma(e)]$}
\end{prooftree}
\end{minipage}
&

\begin{minipage}[h]{.65\linewidth}
\begin{prooftree} 
\AxiomC{$act \equiv \wt{\sigma(e_1)}:\rd{\sigma(\vc{f})}$}
\UnaryInfC{$(x[e_1]:=e,\sigma)\reda{act} \sigma[(x,\sigma(e_1))  \mapsto\sigma(e)]$}
\end{prooftree}
\end{minipage}
\\

\\


\begin{minipage}[h]{0.3\linewidth}
\begin{prooftree} 
\AxiomC{$(c_1,\sigma)\reda{act} \sigma'$}
\UnaryInfC{$( c_1;c_2,\sigma)\reda{act} (c_2, \sigma')$}
\end{prooftree}
\end{minipage}

&

\begin{minipage}[h]{0.65\linewidth}
\begin{prooftree}
\AxiomC{$(c_1,\sigma)\reda{act} (c'_1,\sigma')$}
\UnaryInfC{$( c_1;c_2,\sigma)\reda{act} (c'_1;c_2, \sigma')$}
\end{prooftree}
\end{minipage}

\\

\\

\begin{minipage}[h]{0.3\linewidth}
\begin{prooftree}
\AxiomC{ }
\UnaryInfC{$(\skipp,\sigma)\reda{} \sigma$}
\end{prooftree}
\end{minipage}

&

\begin{minipage}[h]{.65\linewidth}
\begin{prooftree}
\AxiomC{$\sigma(e)  = 1 ~ ? ~ i=1 : \ i=2$}
\AxiomC{$act \equiv \br{\sigma(e)}:\rd{\sigma(\vc{f})}$}
\BinaryInfC{$(\iifthenelse{e}{c_1}{c_2},\sigma)\reda{act} (c_i, \sigma)$}
\end{prooftree}
\end{minipage}
\end{tabular}

\medskip

\begin{tabular}{l}
\begin{minipage}[h]{0.75\linewidth}
\begin{prooftree}
\AxiomC{}
        \UnaryInfC{$(\whilep{e}{c},\sigma)\reda{} (\iifthenelse{e}{c;\whilep{e}{c}}{\skipp},\sigma)$}
\end{prooftree}
\end{minipage}
\end{tabular}
\end{scriptsize}
\caption{Syntax and Labeled Operational semantics}\label{fig:T13}
\end{figure*}

Following \cite{aga2000,almeida2016}, we consider two types of
cache-based information leaks:  $(i)$ disclosures that happen when secret data determine which parts of the
program are executed;  $(ii)$ disclosures that arise when access to memory is indexed by sensitive information. 
In order to model the latter category, we shall enrich the simple language from section \ref{s:2} with {\em arrays}: 
\[
\begin{array}{lll}
c & ::= &x:=e ~|~  x[e_1]:=e ~|~ \skipp~|~ c_1;c_2~|~ \iifthenelse{e}{c_1}{c_2}~|~ \whilep{e}{c}
\end{array}
\]

\begin{figure*}[!tbp]
	\begin{scriptsize}
\begin{tabular}{l}
\begin{minipage}[h]{1.0\linewidth}
\begin{prooftree}
\AxiomC{$x\not\in X_O$} 
   \LeftLabel{As1'}
\UnaryInfC{\ctype{p}{\trH{\Gamma}{x:=e}{\Gamma[x\mapsto p\js \Gamma[\alpha](fv(e))][x_l\mapsto \Gamma(x_l)\js \Gamma[\alpha](fv(\vc{f})))]}}}
\end{prooftree}
\end{minipage}
\\
\\
\begin{minipage}[h]{1.0\linewidth}
\begin{prooftree}
\AxiomC{$x\not\in X_O$} 
\noLine
\UnaryInfC{$p_1=(\Gamma[\alpha](fv(e_1),fv(e))$}
\AxiomC{$p_l=(\Gamma[\alpha](fv(e_1),fv(\vc{f}))$}
   \LeftLabel{Ast'}
\BinaryInfC{\ctype{p}{\trH{\Gamma}{x[e_1]:=e}{\Gamma[x\mapsto p\js \Gamma(x) \js p_1][x_l\mapsto \Gamma(x_l)\js p_l]}}}
\end{prooftree}
\end{minipage}
\\
\\
\begin{minipage}[h]{1.0\linewidth}
\begin{prooftree}
\AxiomC{$x\in X_O\setminus fv(e)$} 
\AxiomC{$\Gamma_1=\Gamma\lhd_\alpha x$}
   \LeftLabel{As2'}
\BinaryInfC{\ctype{p}{\trH{\Gamma}{x:=e}{\Gamma_1[x\mapsto p\js \Gamma_1[\alpha](fv(e))][x_l\mapsto \Gamma_1(x_l)\js \Gamma_1[\alpha](fv(\vc{f})))]}}}
\end{prooftree}
\end{minipage}
\\
\\
\begin{minipage}[h]{1.0\linewidth}
\begin{prooftree}
\AxiomC{$x\in  X_O\cap fv(e)$}
\AxiomC{$\Gamma_1=\Gamma\lhd_\alpha x$}
   \LeftLabel{As3'}
\BinaryInfC{\ctype{p}{\trH{\Gamma}{x:=e}{\Gamma_1[x \mapsto  p\js \Gamma(x) \sqcup \Gamma_1[\alpha](fv(e)\setminus x)][x_l\mapsto \Gamma_1(x_l)\js \Gamma_1[\alpha](fv(\vc{f})))]}}}
\end{prooftree}
\end{minipage}
\\
\\
\begin{minipage}[h]{1.0\linewidth}
\begin{prooftree}
 \AxiomC{}
   \LeftLabel{Skip}
\UnaryInfC{\ctype{p}{\trH{\Gamma}{skip}{\Gamma}}}
\end{prooftree}
\end{minipage}
\\
\\
\begin{minipage}[h]{1.0\linewidth}
\begin{prooftree}
 	\AxiomC{\ctype{p}{\trH{\Gamma}{c_1}{\Gamma_1}}}
\AxiomC{\ctype{p}{\trH{\Gamma_1}{c_2}{\Gamma_2}}}
   \LeftLabel{Seq}
\BinaryInfC{\ctype{p}{\trH{\Gamma}{c_1;c_2}{\Gamma_2}}}
\end{prooftree}
\end{minipage}

\\
\\
\begin{minipage}[h]{1.0\linewidth}
\begin{prooftree}
	\AxiomC{$p_0 \sqsubseteq_r p_1$\hspace*{-2em}}
 \AxiomC{$\Gamma \sqsubseteq_r \Gamma'$\hspace*{-2em}}
 \AxiomC{\ctype{p_1}{\trH{\Gamma'}{c}{\Gamma_1'}}\hspace*{-2em}}
\AxiomC{$\Gamma_1'\sqsubseteq_r \Gamma_1$}
 \LeftLabel{Sub}
\QuaternaryInfC{\ctype{p_0}{\trH{\Gamma}{c}{\Gamma_1}}}
\end{prooftree}
\end{minipage}
\\
\\

\begin{minipage}[h]{1.0\linewidth}
\begin{prooftree}
\AxiomC{$p_l=\Gamma[\alpha](fv(e))$}
\noLine
\UnaryInfC{$p'=(p_l, \Gamma) \lhd_\alpha (\affo(c_1) \cup \affo(c_2)) $}
\AxiomC{\ctype{p\js p'}{\trH{\Gamma[x_l\mapsto \Gamma(x_l)\js p_l]}{c_i}{\Gamma_i}}}
\noLine
\UnaryInfC{$\Gamma'=  \Gamma_1 \lhd_\alpha \affo(c_2)\js \Gamma_2 \lhd_\alpha \affo(c_1)$}
  \LeftLabel{If}
\BinaryInfC{\ctype{p}{\trH{\Gamma}{\iifthenelse{e}{c_1}{c_2}}{\Gamma'}}}
\end{prooftree}
\end{minipage}
\\

\\
\begin{minipage}[h]{1.0\linewidth}
\begin{prooftree}
\AxiomC{$p_l=\Gamma[\alpha](fv(e))$}
\noLine
\UnaryInfC{$p_e=\Gamma_1[\alpha](fv(e))$}
\AxiomC{\ctype{p\js p_e}{\trH{\Gamma_1[x_l\mapsto \Gamma(x_l)\js p_e]}{c}{\Gamma'}}}
\noLine
\UnaryInfC{$\Gamma[x_l\mapsto \Gamma(x_l)\js p_l] \js ((\Gamma'[x_l\mapsto \Gamma'(x_l)\js p_e]) \lhd_\alpha \affo(c)) \sbs_r \Gamma_1$}
 \LeftLabel{Wh}
\BinaryInfC{\ctype{p}{\trH{\Gamma}{\whilep{e}{c}}{\Gamma_1}}}
\end{prooftree}
\end{minipage}

\end{tabular}
\comm{
\begin{tabular}{c}
\\

\begin{minipage}[h]{\linewidth}
\begin{prooftree}
\AxiomC{\nntype{p\js p_e}{\trH{\Gamma_1}{c}{\Gamma'}}}
\AxiomC{$p_e=\Gamma_1[\alpha](fv(e))$}
\noLine
\UnaryInfC{$(\Gamma \lhd_\alpha \affo(c))\js (\Gamma' \lhd_\alpha \affo(c)) \sbs_r \Gamma_1$}
 \LeftLabel{Wh}
\BinaryInfC{\nntype{p}{\trH{\Gamma}{\whilep{e}{c}}{\Gamma_1}}}
\end{prooftree}
\end{minipage}
\end{tabular}
}
	\end{scriptsize}
\caption{Typing Rules for Output Sensitive Constant Time}\label{fig:T14}
\end{figure*}

To simplify notations, we assume that array indexes $e_1$ are basic expressions (not referring to arrays) and that $X_O$ 
does not contain arrays.   
Moreover as in \cite{almeida2016}, 
a state or store $\sigma$ maps array variables $v$ and indices $i\in\nat$ to values $\sigma(v,i)$.
The labeled semantics of While programs is listed in Figure \ref{fig:T13}. 
In all rules, we denote $\vc{f}=(f_i)_i$, the set of all indexes occurring in $e$
(i.e., $e$ contains sub-expressions of the form  $x_i[f_i]$). 

The labels on the execution steps correspond to the information which is leaked to the environment ($\rd{}$ for a read access on memory, $\wt{}$ for a write access and $\br{}$ for a branch operation). In the rule for (If), the valuations of branch conditions are leaked. Also, all indexes to program variables read and written at each statement are leaked to. Remark that in cache based attacks, only the offsets are leaked and not the base variable addresses/values. When there is no label on a step, this step is considered to be invisible.

We give in Figure~\ref{fig:T14}  the new typing rules. As above, we denote $\vc{f}=(f_i)_i$,  the set of all indexes occurring in $e$. We add a fresh variable $x_l$,
that is not used in programs, in order to capture the unintended leakage. Its type is always growing up and it mirrors the information leaked by each command.
In rule $(As1")$ we take a conservative approach and we consider that the type of an array variable is the lub of all its cells. The information leaked by the assignment $x[e_1]:=e$ is the index $e_1$ plus the set $\vc{f}=(f_i)_i$ of all indexes occurring in $e$. Moreover, the new type of the array variable $x$ mirrors the fact that now the value of $x$ depends also on the index $e_1$ and on the right-hand side $e$. 

\begin{defi}\label{d3.2}
An  {\bf execution trace} (or simply trace)  is a sequence of visible actions $\reda{a_1}\reda{a_2}\ldots \reda{a_n}$, all invisible steps being ignored.
A program $c$ is {\bf  ($X_I, X_O$)-constant time} when all its traces starting with $X_I $-equivalent stores that lead to finally $X_O$-equivalent stores,  are identical.
\end{defi}
Following \cite{almeida2016}, given a set $X$ of program variables, two stores $\sigma$ and $\rho$  are {\it $X$-equivalent} when $\sigma(x, i)=\rho(x, i)$ for all $x\in X$ and $i\in\nat$.
Two traces  $\reda{a_1}\ldots \reda{a_n}$ and  $\reda{b_1}\ldots \reda{b_m}$ are {\it identical} iff $n=m$ and $a_j=b_j$ for all $1\leq j \leq n$. 

We can reduce the ($X_I, X_O$)-constant time security of a command $c$ to the ($X_I, X_O, \{x_l\}$)-security (see section \ref{s:iol}) of a corresponding command $\omi(c)$, obtained by adding a fresh variable $x_l$ to the program variables $fv(c)$, and then adding recursively before each assignment and each boolean condition predicate, a new assignment to the leakage variable $x_l$ that mirrors the leaked information. Let $:, \br, \rd, \wt$ be some new abstract operators. The construction of the instrumentation $\omi(\bullet)$ is shown in Fig. \ref{fig:ga}.  As above, we denote by $\vc{f}=(f_i)_i$ the set of all indexes occurring in $e$. 

First we can extend the \While\ language with array variables, then we need to extend the typing system from section \ref{s:2} with a rule corresponding to the new rule $Ast'$.
Then, the following lemma gives the relationship between the type of a program $c$ using  the new typing system and the type of the instrumented program $\omi(c)$ using  the extended typing system from the previous section. 

\begin{figure*}[!tbp]
\[
\begin{array}{l|l}
\bullet  &  \omi(\bullet) \\
\hline
x:=e & x_l:=x_l:\rd{\vc{f}}; \ \  x:=e \\
\hline
x[e_1]:=e & x_l:=x_l:\wt{e_1}:\rd{\vc{f}}; \ \  x[e_1]:=e\\
\hline
skip & skip \\
\hline
c_1;c_2 & \omi(c_1);\omi(c_2) \\
\hline
\iifthenelse{e}{c_1}{c_2}  & x_l:=x_l:\br{e}:\rd{\vc{f}}; \ \ \iifthenelse{e}{\omi(c_1)}{\omi(c_2)}\\
\hline
\whilep{e}{c}  &x_l:=x_l:\br{e}:\rd{\vc{f}}; \textsf{While } e \textsf{ Do } \omi(c);x_l:=x_l:\br{e}:\rd{\vc{f}} \textsf{ oD}\\
\end{array}
\]
\caption{Instrumentation for $\omi(\bullet)$}\label{fig:ga}
\end{figure*}
\begin{lem}\label{al:ct}
Let $c$ a command such that $x_l\not\in fv(c)$,  $\sigma, \sigma'$ two stores , $tr$ some execution trace and $[]$ the empty trace. 

\noindent{\it 1.\ } \ctype{p}{\trH{\Gamma}{c}{\Gamma'}} iff  \nntype{p}{\trH{\Gamma}{\omi(c)}{\Gamma'}}.
\item  

\noindent{\it 2.\ } $(c,\sigma)\stackrel{tr}{\Longrightarrow} \sigma'$ iff  $(\omi(c),\sigma[x_l\mapsto []])\Longrightarrow \sigma'[x_l\mapsto tr]$.
\end{lem}
\comm{
\begin{proof}
By structural induction on $c$.
\begin{itemize}
\item $c$ is an assignment $x:=e$ for some $x\not\in X_O$.\\ 
Then $\omi(c) \deff x_l:=x_l:\rd{\vc{f}}; \ \  x:=e$.
On one side, using twice the rule (As1), we get \nntype{p}{\trH{\Gamma}{x_l:=x_l : \rd{\vc{f}}}{\Gamma[x_l\mapsto \Gamma(x_l)\js \Gamma[\alpha](fv(\vc{f})))]}} and  \nntype{p}{\trH{\Gamma[x_l\mapsto \Gamma(x_l)\js \Gamma[\alpha](fv(\vc{f})))]}{x:=e}{\Gamma[x\mapsto p\js \Gamma[\alpha](fv(e))][x_l\mapsto \Gamma(x_l)\js \Gamma[\alpha](fv(\vc{f})))]}} and then we use the rule (Seq) to get
 \nntype{p}{\trH{\Gamma}{\omi(c)}{\Gamma[x\mapsto p\js \Gamma[\alpha](fv(e))][x_l\mapsto \Gamma(x_l)\js \Gamma[\alpha](fv(\vc{f})))]}}.
On the other side we apply once the rule (As1') and we get \ctype{p}{\trH{\Gamma}{x:=e}{\Gamma[x\mapsto p\js \Gamma[\alpha](fv(e))][x_l\mapsto \Gamma(x_l)\js \Gamma[\alpha](fv(\vc{f})))]}}.
\qedhere

\end{itemize}
\end{proof}
}
Now combining Theorem \ref{l:sec} and Lemma \ref{al:ct} we get the following Theorem which proves the soundness of the new typing system.
\begin{thm}\label{l:ctime}
Let $\cL$  be the lattice defined in the section \ref{s:iol}. 
Let $(\Gamma,\alpha)$ be defined by $\Gamma(x)=\{\tau_x\}$, for all $x\in Var$ and  $\alpha(o)=\{\tau_{\overline{o}}\}$, for all $o\in X_O$ and $\Gamma(x_l)=\bot$.   If \ctype{p}{\trH{\Gamma}{c}{\Gamma'}}  and  $\Gamma'(x_l) \sbs \Gamma(X_I) \js \alpha(X_O)$, then $c$ is ($X_I, X_O$)- constant time.
\end{thm}

\section{Application to low-level code}\label{llvm:sec}

We show in this section how the type system we proposed to express output-sensitive constant-time non-interference 
on the {\em While} language can be lifted to a low-level program representation like the LLVM byte code~\cite{lattner2004}.
\subsection{LLVM-IR}
\begin{figure*}[!tbp]
\begin{center}
\begin{tabular}{|l|l|} \hline 
	$r \leftarrow op(Op, \overrightarrow{v})$ & assign to $r$ the result of $Op$ applied to operands $ \overrightarrow{v}$  \\ \hline 
	$r \leftarrow load(v)$ &  load in $r$ the value stored at address pointed by $v$ \\ \hline 
	$store (v_1, v_2)$  & store at address pointed by $v_2$ the value   $v_1$ \\ \hline 
	$cond (r, b_{then}, b_{else})$  & branch to $b_{then}$ if the value of $r$ is true and to $b_{false}$ otherwise  \\ \hline 
	$goto~b$ & branch to $b$ \\ \hline
\end{tabular}
\end{center}
\caption{Syntax and informal semantics of simplified LLVM-IR} 
\label{fig:tab-llvm}
\end{figure*}
We consider a simplified LLVM-IR representation with four instructions: assignments from an expression (register or immediate value) or from a memory block (load), writing to a memory block (store) and 
(un)conditional jump instructions. We assume that the program control flow is represented by a control-flow graph (CFG) $G = (\cB, \rightarrow_E, b_{init}, b_{end})$ where $\cB$ is the 
set of basic blocks, $\rightarrow_E$ the set of edges connecting the basic blocks, $b_{init} \in \cB$ the entry point and $b_{end} \in \cB$ the ending point.
We denote by $Reach(b, b')$ the predicate indicating that  node $b'$ is {\em reachable} from node $b$, i.e., there exists a path in $G$ from $b$ to $b'$. 
A program is then a (partial) map from control points $(b,n) \in \cB \times \mathbb{N}$ to instructions where 
each basic block is terminated by a jump instruction.
The memory model consists in a set of {\em registers or temporary variables} $R$ and 
a set of  memory blocks $M$ (including the execution stack). $Val$ is the set of values and memory block addresses.
The informal semantics of our simplified LLVM-IR is given in Figure~\ref{fig:tab-llvm}, where $r \in R$ and $v \in R \cup Val$ is a register or an immediate value. 

In the formal operational semantics, execution steps are labeled with leaking data, i.e., 
addresses of store and load operations and branching conditions. This formal  semantics is defined in Figure~\ref{fig:sem-llvm}.
It is implicitly parameterized by the program $p$, a configuration is a tuple $((b,n), \rho, \mu)$ where $(b,n) \in \cB \times \mathbb{N}$ is 
the control point and $\rho: R \rightarrow Val$ (\emph{resp.} $\mu: M \rightarrow Val$) denotes the content of registers (\emph{resp.} memory).

\begin{figure*}[!htbp]
\begin{center}
\begin{tabular}{l}
\begin{minipage}[h]{\linewidth}
\begin{prooftree}
\AxiomC{$p(b,n) = r \leftarrow op(Op, \overrightarrow{v})$}
\AxiomC{$v = \overline{Op}~(\llbracket \overrightarrow{v} \rrbracket_{(\rho, \mu)})$}
\BinaryInfC{$((b,n), \rho, \mu) \longrightarrow ((b,n+1), \rho[r \rightarrow v], \mu)$} 
\end{prooftree}
\end{minipage}

\\

\\

\begin{minipage}[h]{\linewidth}
\begin{prooftree}
\AxiomC{$p(b,n) = r \leftarrow load(v)$}
\AxiomC{$ad = \llbracket v \rrbracket_{(\rho, \mu)}$}
\AxiomC{$val = \mu(ad)$}
\TrinaryInfC{$((b,n), \rho, \mu) \reda{\rd{ad}} ((b,n+1), \rho[r \rightarrow val], \mu)$} 
\end{prooftree}
\end{minipage}

\\
\\

\begin{minipage}[h]{\linewidth}
\begin{prooftree}
\AxiomC{$p(b,n) = goto~b'$}
\UnaryInfC{$((b,n), \rho, \mu) \longrightarrow ((b',0), \rho, \mu)$} 
\end{prooftree}
\end{minipage}

\\
\\
\begin{minipage}[h]{\linewidth}
\begin{prooftree}
\AxiomC{$p(b,n) = store(v_1, v_2)$}
\AxiomC{$v = \llbracket v_1 \rrbracket_{(\rho, \mu)}$}
\AxiomC{$ad = \llbracket v_2 \rrbracket_{(\rho, \mu)}$}
\TrinaryInfC{$((b,n), \rho, \mu) \reda{\wt{ad}} ((b,n+1), \rho, \mu[ad \rightarrow v])$} 
\end{prooftree}
\end{minipage}


\\
\\

\begin{minipage}[h]{\linewidth}
\begin{prooftree}
\AxiomC{$p(b,n) = cond (r, b_{then}, b_{else})$}
\AxiomC{$\llbracket r \rrbracket_{(\rho, \mu)} = 1$}
\BinaryInfC{$((b,n), \rho, \mu)  \reda{\ju{1}}  ((b_{then},0), \rho, \mu)$} 
\end{prooftree}
\end{minipage}

\\
\\


\begin{minipage}[h]{\linewidth}
\begin{prooftree}
\AxiomC{$p(b,n) = cond (r, b_{then}, b_{else})$}
\AxiomC{$\llbracket r \rrbracket_{(\rho, \mu)} = 0$}
\BinaryInfC{$((b,n), \rho, \mu)  \reda{\ju{0}} ((b_{else},0), \rho, \mu)$} 
\end{prooftree}
\end{minipage}
\end{tabular}
\end{center}
\caption{Labeled operational semantics for LLVM-IR}
\label{fig:sem-llvm}
\end{figure*}


\subsection{Type system}\label{sec:type-system-llvm}

First, we introduce the following notations for an LLVM-IR program represented by a  CFG $G = (\cB, \rightarrow_E, b_{init}, b_{end})$: 
\begin{enumerate}

\item  Function $dep: \cB \rightarrow 2^\cB$ associates to each basic block its set of ``depending blocks'', i.e., 
$b' \in dep(b)$ iff $b'$ dominates $b$ and there is no block $b"$ between $b'$ and $b$ such that $b"$ post-dominates $b'$. 
We recall that a node  $b_1$ dominates ({\em resp.} post-dominates) a node  $b_2$ iff every path from the entry node  $b_{init}$ to $b_2$ goes through $b_1$ 
({\em resp.} every path from $b_2$ to the ending node $b_{end}$ goes through $b_1$).

\item Partial function $br: \cB \rightarrow R$ returns the ``branching register'', i.e., the register $r$ used to compute 
the branching condition leading outside $b$ ($b$ is terminated by an instruction $cond (r, b_{then}, b_{else})$).
Note that in LLVM branching registers are always {\em fresh} and assigned only once before to be used.

\item Function $PtsTo: (\cB \times \mathbb{N}) \times (R\cup M) \rightarrow  2^M$ returns the set of (addresses of) memory blocks pointed to by a given register or memory block  at a given control point. 
For example,  $bl \in PtsTo(b,n)(r)$ means that at control point $(b,n)$, 
register $r$  may contain the address of block $bl\in M$.\medskip
\end{enumerate}

We now define a type system (Figures~\ref{fig:llvm-op} to \ref{fig:llvm-jump}) that allows to express the  output-sensitive constant-time property for LLVM-IR -like programs.
The main difference with respect to the rules given at the source level (Figures~\ref{fig:T1} and \ref{fig:T14}) 
is that the control-flow is explicitly given by the CFG, and not by the language syntax.  For a LLVM-like program, an environment $\Gamma : R \cup M \mapsto \cL$, associates security types to registers and memory blocks.  
We will use the notation 
\mbox{$\vdash_\alpha (b,n) : \Gamma  \Rightarrow \Gamma'$}
inspired from the one used by~\cite{bar2014}. The intuitive meaning of this judgement is the following: if $I$ is the instruction of control point $(b,n)$ and $\tau_0$ is
the join of all the types of the branching conditions dominating the current basic block $b$  then  $\vdash_\alpha (b,n) : \Gamma  \Rightarrow \Gamma'$ is equivalent to the notation \nntype{\tau_0}{\trH{\Gamma}{I}{\Gamma'}} used in section~\ref{sec-typing-rules}, that is, it transforms the previous environment $\Gamma$ into the new environment $\Gamma'$.


\begin{figure*}[!htb]
	\begin{center}
		\begin{tabular}{l}
\begin{minipage}[h]{\linewidth}
\begin{prooftree}
\AxiomC{$p(b,n) = r \leftarrow op(Op, \overrightarrow{v})$}
\noLine
\UnaryInfC{$\tau_0 = \displaystyle\bigsqcup_{x \in br(dep(b))} \Gamma[\alpha](x)$}
\AxiomC{$\tau = (\Gamma[\alpha](\overrightarrow{v})\;\; \sqcup \tau_0)$} 
\noLine
\UnaryInfC{$r \not\in X_O$}
\LeftLabel{Op1}
\BinaryInfC{$\vdash_\alpha (b,n) : \Gamma  \Rightarrow \Gamma[r \rightarrow \tau] $}
\end{prooftree}
\end{minipage}
\\
\\
\begin{minipage}[h]{\linewidth}
\begin{prooftree}
\AxiomC{$p(b,n) = r \leftarrow op(Op, \overrightarrow{v})$}
\noLine
\UnaryInfC{$\tau_0 = \displaystyle\bigsqcup_{x \in br(dep(b))} \Gamma[\alpha](x)$}
\AxiomC{$\tau = (\Gamma[\alpha](\overrightarrow{v})\;\; \sqcup \tau_0)$} 
\noLine
\UnaryInfC{$r \in X_O$}
\noLine
\UnaryInfC{$\Gamma_1 = \Gamma \lhd_\alpha r$}
\LeftLabel{Op2}
\BinaryInfC{$\vdash_\alpha (b,n) : \Gamma \Rightarrow \Gamma_1[r \rightarrow \tau]$}
\end{prooftree}
\end{minipage}
\end{tabular}
	\end{center}
	\caption{assignment from an operation $Op$}
	\label{fig:llvm-op}
\end{figure*}
\paragraph{\bf Assignment from an operation $Op$ (Figure~\ref{fig:llvm-op}).} In rules Op1 and Op2, 
the new type $\tau$ of the assigned register $r$ 
is the join of the type of operands $\overrightarrow{v}$ and the type of all the branching conditions dominating the current basic block ($\tau_0$). 
Note that since branching registers $r$ are assigned only once in LLVM there is no need to update their dependencies from output variables (using the $\lhd_\alpha$ operator),
$\Gamma(r)$ being never changed once $r$ has been assigned.
\begin{figure*}[!htb]
	\begin{center}
\begin{minipage}[h]{\linewidth}
\begin{prooftree}
\AxiomC{$p(b,n) = r \leftarrow load(v)$}
\noLine
\UnaryInfC{$\tau_0 = \displaystyle\bigsqcup_{x \in br(dep(b))} \Gamma[\alpha](x)$}
\AxiomC{$A_m = PtsTo (b,n)(v)$}
\noLine
\UnaryInfC{$\tau_1 = \Gamma[\alpha](v) \sqcup \tau_0$}
\AxiomC{$ r \not\in X_O$}
\noLine
\UnaryInfC{$\tau_2 = \Gamma[\alpha](A_m)$}
\LeftLabel{Ld1}
\TrinaryInfC{$\vdash_\alpha (b,n) : \Gamma \Rightarrow (
\Gamma[x_l \rightarrow \Gamma(x_l) \sqcup \tau_1][r \rightarrow \tau_2 \sqcup \tau_1]$}
\end{prooftree}
\end{minipage}

~ \\
~ \\
~ \\

\begin{minipage}[h]{\linewidth}
\begin{prooftree}
\AxiomC{$p(b,n) = r \leftarrow load(v)$}
\noLine
\UnaryInfC{$\tau_0 = \displaystyle\bigsqcup_{x \in br(dep(b))} \Gamma_1[\alpha](x)$}
\AxiomC{$A_m = PtsTo (b,n)(v)$}
\noLine
\UnaryInfC{$\tau_1 = \Gamma_1[\alpha](v) \sqcup \tau_0$}
\AxiomC{$r \in X_O$}
\noLine
\UnaryInfC{$\Gamma_1 = \Gamma \lhd_\alpha r$}
\noLine
\UnaryInfC{$\tau_2 = \Gamma_1[\alpha](A_m)$}
\LeftLabel{Ld2}
\TrinaryInfC{$\vdash_\alpha (b,n) : \Gamma \Rightarrow 
\Gamma_1[x_l \rightarrow \Gamma_1(x_l) \sqcup \tau_1][r \rightarrow \tau_2 \sqcup \tau_1]$}
\end{prooftree}
\end{minipage}
	\end{center}
	\caption{assignment from a load expression}
	\label{fig:llvm-load}
\end{figure*}
\paragraph{\bf  Assignment from a load expression (Figure~\ref{fig:llvm-load}).}
Rules Ld1 and Ld2 update $\Gamma$ in a similar way as Op1 and Op2, the main difference being that  since some of the memory locations accessed when dereferencing $v$ (i.e., $PtsTo(b,n)(v)$) are in $A_m$ 
(i.e., potentially in the cache) the dependencies of $v$ are added to the type of the leakage variable $x_l$.
\begin{figure*}[!htb]
	\begin{center}
\begin{minipage}[h]{\linewidth}
\begin{prooftree}
\AxiomC{$p(b,n) = store(v_1, v_2)$}
\noLine
\UnaryInfC{$\tau_0 = \displaystyle\bigsqcup_{x \in br(dep(b))} \Gamma[\alpha](x)$}
\AxiomC{$A_m = PtsTo (b,n)(v_2)$}
\noLine
\UnaryInfC{$A_0 = A_m \cap X_0$}
\noLine
\UnaryInfC{$\tau_1 = \Gamma_1[\alpha](v_2) \sqcup \tau_0$}
\AxiomC{$\Gamma_1 = \Gamma \lhd_\alpha A_0$}
\noLine
\UnaryInfC{$\tau_2 = \Gamma_1[\alpha](v_1)$}
\LeftLabel{St}
\TrinaryInfC{$\vdash_\alpha (b,n) : \Gamma \Rightarrow	\Gamma_1[x_l \rightarrow \Gamma_1(x_l) \sqcup \tau_1][v_{s_{\in A_m}} \rightarrow \Gamma(v_s) \sqcup \tau_2 \sqcup \tau_1]$}
\end{prooftree}
\end{minipage}
	\end{center}
	\caption{store instruction}
	\label{fig:llvm-store}
\end{figure*}
\paragraph{\bf  Store instruction (Figure~\ref{fig:llvm-store}).} 
Rule St  updates the dependencies of all memory locations pointed to by $v_2$  by adding the types of $v_1$ and $v_2$ itself. In addition, the type of the leakage variable $x_l$ is also updated with the dependencies of  $v_2$ and with the dependencies of all branching registers that influenced the execution flow to reach the current block $b$.

\begin{figure*}[!htb]
	\begin{center}
	\begin{tabular}{lr}
\begin{minipage}[h]{\linewidth}
\begin{prooftree}
\AxiomC{$p(b,n) = cond (r, b_{then}, b_{else})$}
\LeftLabel{CJmp}
\UnaryInfC{$\vdash_\alpha (b,n) : \Gamma \Rightarrow 
\Gamma[x_l \rightarrow \Gamma(x_l) \sqcup \Gamma[\alpha](r)]$}
\end{prooftree}
\end{minipage}

&

\begin{minipage}[h]{\linewidth}
\begin{prooftree}
\AxiomC{$p(b,n) = goto~b'$}
\LeftLabel{Jmp}
\UnaryInfC{$\vdash_\alpha (b,n) : \Gamma \Rightarrow \Gamma$}
\end{prooftree}
\end{minipage}
\end{tabular}
	\end{center}
	\caption{conditional and unconditional jump}
	\label{fig:llvm-jump}
\end{figure*}
\paragraph{\bf  Conditional and unconditional jump (Figure~\ref{fig:llvm-jump}).} 
Rule CJmp indicates that the leakage variable type is augmented with the type of the branching condition register. 
Unconditional jumps (Rule Jmp) leave the environment unchanged.

\subsection{Well typed LLVM programs are  output-sensitive constant-time}

The following definition is adapted from~\cite{bar2014}.

\begin{defi}\label{wt0:llvm}
	An LLVM-IR  program $p$ is well typed with respect to an initial environment $\Gamma_0$ and final environment $\Gamma'$ (written $\vdash_\alpha p : \Gamma_0  \Rightarrow \Gamma'$) , if there is a family of {\em well-defined} environments  $\{(\Gamma)_{(b,n)} \ \mid \ (b,n)\in (\cB, \nat)\}$, such that for all nodes $(b,n)$ and all its successors $(b',n')$, there exists a type environment $\gami$ and $A\subseteq X_O$ such that  
\[\vdash_\alpha (b,n) : \Gamma_{(b,n)} \Rightarrow \gami \text{\ and \ }  (\gami\lhd_\alpha A)  \sbs_r \Gamma_{(b',n')}.\]
\end{defi}
In the above definition the set $A$ is {\bf mandatory } in order to prevent dependency cycles between variables in $X_O$.

The following Theorem is the counterpart of Theorem~\ref{l:ctime}. It shows the soundness of our type system for LLVM-IR programs with respect to output-sensitive constant-time.

\begin{thm}\label{llvm:ctime}
Let $\cL$  be the lattice from the section \ref{s:iol}. 
Let $(\Gamma,\alpha)$ be defined by $\Gamma(x)=\{\tau_x\}$, for all $x\in R\cup M$,  $\alpha(o)=\{\tau_{\overline{o}}\}$, for all $o\in X_O$ and $\Gamma(x_l)=\bot$.   If \ $\vdash_\alpha p : \Gamma \Rightarrow \Gamma' $ and  $\Gamma'(x_l) \sbs \Gamma(X_I) \js \alpha(X_O)$, then $p$ is ($X_I, X_O$)- constant time.
\end{thm}

\subsection{Example}

We illustrate below the effect of the LLVM-IR typing rules  on a short example.
The C code of this example is given on Figure~\ref{fig:c-ex}, and the corresponding (simplified) LLVM-IR on Figure~\ref{fig:llvm-ex}.

\begin{figure*}[!htb]
\begin{center}
\begin{lstlisting}[language=C]
int p[10], q[10] ; // global variables

int main () {
   int x, y ;
   p[x] = q[y] ;
   return 0 ;      // output value is always 0
}
\end{lstlisting}
	\caption{A C code example}
	\label{fig:c-ex}
\end{center}
\end{figure*}

\begin{figure*}[!htb]
\begin{center}
\begin{lstlisting}[language=LLVM, style=numbered]
  @q = common global                   
  @p = common global                   
  %x = alloca i32                      
  %y = alloca i32                      

  %1 = load i32* %y                   
  %2 = getelementptr @q, 0, %1       
  %3 = load %2                      
  %4 = load %x                     
  %5 = getelementptr @p, 0, %4    
  store %3, %5                   
\end{lstlisting}
\caption{The simplified LLVM-IR code of Figure~\ref{fig:c-ex}}
	\label{fig:llvm-ex}
\end{center}
\end{figure*}

In LLVM-IR, instructions {\tt alloca} and {\tt global} allow to declare temporary and global variables and their value correspond to memory addresses.
For a sake of simplicity we choose to not define specific typing rules for these instructions, they 
are taken into account only in building the initial environment. 

First, we assume that $x_l$ denotes the \emph{leakage} variable and that the content of C variables {\tt p}, {\tt q}, {\tt x} and {\tt y} are stored in memory blocks $b_0$ to $b_3$, i.e. at the initial control point, $PtsTo(@p)=b_0, PtsTo(@q)=b_1, PtsTo(\%x)=b_2, PtsTo(\%y)=b_3$.

Now we consider the following initial environment (produced by lines 1-4 in Figure~\ref{fig:llvm-ex}):  

\noindent
$\Gamma_0(@p) = \Gamma_0(@q) = \Gamma_0(\%x) = \Gamma_0(\%y) = \bot $ \\
$\Gamma_0(b_0) = P, \Gamma_0(b_1) = Q, \Gamma_0(b_2) = X, \Gamma_0(b_3) = Y, \Gamma_0(x_l) = \bot$

This  initial environment captures the idea that the values of variables $@p$, $@q$, $\%x$, $\%y$ are addresses (of memory blocks corresponding to the ``high-level''  C variables $p$, $q$, $x$ and $y$) and hence their security type is $\bot$, and the  memory blocks $b_0$ to $b_3$ correspond to the C variables $p$, $q$, $x$ and $y$, and this is mirrored in the initial  environment $\Gamma_0$.  Moreover, initially, nothing is leaked yet.

We then update $\Gamma_0$ by applying our typing rules in sequence to each instruction of the LLVM-IR representation. Note that the {\tt getelementptr} instruction, which is specific to LLVM, allows to compute an address 
corresponding to an indexed access in a buffer. Hence, it is treated by our typing system as an arithmetic ($Op$) instruction.

\vspace*{1em}
\noindent
  {\tt \%1 = load \%y}
\begin{eqnarray*}
	\Gamma_1 & = & \Gamma_0[x_l \rightarrow \Gamma_0(x_l) \sqcup \Gamma_0(\%y), \%1 \rightarrow \Gamma_0(\%y) \sqcup \Gamma_0(PtsTo(\%y))] \\
		& = & \Gamma_0[x_l \rightarrow \bot, \%1 \rightarrow Y]
\end{eqnarray*}

\noindent
  {\tt \%2 = getelementptr @q, 0, \%1}
\begin{eqnarray*}
	PtsTo(\%2)& =& b_1 \\
	\Gamma_2 & = & \Gamma_1[\%2 \rightarrow \Gamma_1(@q) \sqcup  \Gamma_1(0) \sqcup \Gamma_1(\%1)] \\
	& = & \Gamma_1[\%2 \rightarrow Y]
\end{eqnarray*}

\noindent
  {\tt \%3 = load \%2}
\begin{eqnarray*}
	\Gamma_3 & = & \Gamma_2[x_l \rightarrow \Gamma_2(x_l) \sqcup \Gamma_2(\%2), \%3 \rightarrow \Gamma_0(\%2) \sqcup \Gamma_0(PtsTo(\%2))] \\
	& = & \Gamma_2[x_l \rightarrow Y, \%3 \rightarrow Y \js Q]	
\end{eqnarray*}

\noindent
  {\tt \%4 = load \%x}
\begin{eqnarray*}
	\Gamma_4 & = & \Gamma_3[x_l \rightarrow \Gamma_3(x_l) \sqcup \Gamma_3(\%x), \%4 \rightarrow \Gamma_3(\%x) \sqcup \Gamma_3(PtsTo(\%x))]  \\
	         & = & \Gamma_3[\%4 \rightarrow X]
\end{eqnarray*}

\noindent
  {\tt \%5 = getelementptr @p, 0, \%4}
\begin{eqnarray*}
	PtsTo(\%5)& =& b_0 \\
	\Gamma_5 & = & \Gamma_4[\%5 \rightarrow \Gamma_4(@p) \sqcup  \Gamma_4(0) \sqcup \Gamma_4(\%4)] \\
	& = & \Gamma_4[\%5 \rightarrow X]
\end{eqnarray*}
  
\noindent
  {\tt store \%3, \%5}
\begin{eqnarray*}
	\Gamma_6 & = & \Gamma_5[x_l \rightarrow \Gamma_5(x_l) \sqcup \Gamma_5(\%5), PtsTo(\%5) \rightarrow \Gamma_5(PtsTo(\%5)) \sqcup \Gamma_5(\%3) \sqcup \Gamma_5(\%5)] \\
	& = & \Gamma_5[x_l \rightarrow Y \sqcup X, b_0 \rightarrow P \js Y \js Q \sqcup X]
\end{eqnarray*}

Making all the replacements, we get that the final environment is:
\begin{eqnarray*}
	\Gamma_6 & = &[x_l \rightarrow Y \sqcup X, b_0 \rightarrow P \js Y \js Q \sqcup X, b_1 \rightarrow Q,  b_2 \rightarrow X, b_3 \rightarrow Y,  @p \rightarrow \bot,  @q \rightarrow \bot,\\
                 &   & \%x \rightarrow \bot,  \%y \rightarrow \bot,  \%1 \rightarrow Y, \%2 \rightarrow Y, \%3 \rightarrow Y\js Q,  \%4 \rightarrow  X,  \%5 \rightarrow  X]  
\end{eqnarray*}
In this final environment $\Gamma_6$, variable $x_l$ depends on the initial types $X$  and $Y$ assigned to memory blocks $b_2$ and $b_3$. This means that  the addresses accessed when reading (resp. writing) 
buffer {\tt p} (resp. {\tt q}) may \emph{leak} to an attacker. 
Hence, if one of the variables  $x$ or $y$ is a secret,  since neither $x$ nor $y$ is an output value, then this program is not output sensitive constant-time, which may lead to a security issue.
\subsection{Implementation}
We are developing a prototype tool implementing the type system 
for  LLVM programs.
This type system consists in computing flow-sensitive dependency relations between program variables. Def.~\ref{wt0:llvm} provides the necessary conditions under which the obtained result is sound (Theorem~\ref{llvm:ctime}). 
We give some technical indications regarding our implementation.

Output variables $X_O$ are defined as function return values and global variables; we do not currently consider arrays nor pointers in $X_O$. 
Control dependencies cannot be deduced from the syntactic LLVM level, we need to 
explicitly compute the dominance relation between basic blocks of the CFG (the $dep$ function).   
Def.~\ref{wt0:llvm} requires the construction of a set $A \subseteq X_O$ to update the environment 
produced at each control locations in order to avoid circular dependencies (when output variable are 
assigned in {\em alternative} execution paths). To identify the set of basic blocks belonging to such alternative 
execution paths leading to a given block, we use the notion of {\em Hammock regions}~\cite{ferrante1987}. 
More precisely, we compute function $Reg: (\cB \times \cB \times (\rightarrow_E)) \rightarrow 2^\cB$, returning the set of {\em Hammock regions} between a basic block 
$b$ and its immediate dominator $b'$ with respect to an incoming edge $e_i$ of $b$. 
Thus, $Reg(b', b, (c,b))$ is the set of blocks belonging to CFG paths going from $b'$ to $b$ without reaching edge $e_i=(c,b)$:
$Reg(b', b, (c,b))  = \{ b_i \mid b' \rightarrow_E b_1 \dots \rightarrow_E b_n \rightarrow_E b \wedge \forall i \in [1,n-1].\; \neg Reach (b_i, c) \}.$
Fix-point computations are implemented using Kildall's algorithm. 
To better handle real-life examples we are currently implementing the $PtsTo$ function, an inter-procedural analysis, and a more precise 
type analysis combining both over- and under-approximations of variable dependencies (see section~\ref{concl:sec}).

\section{Related Work}

\paragraph{{\bf Information flow.}}  There is a large number of papers
on language-based security aiming to prevent undesired information
flows using type systems (see \cite{sab2003}). An information-flow
security type system statically ensures noninterference, i.e. that
sensitive data may not flow directly or indirectly to public channels
\cite{vol1996,mye1999,vau2007,swa2010}.  The typing system presented
in section \ref{s:2} builds on ideas from Hunt and Sands'

As attractive as it is, noninterference is too strict to be useful in
practice, as it prevents confidential data to have any influence on
observable, public output: even a simple password checker function
violates noninterference.  Relaxed definitions of noninterference have
been defined in order to support such intentional downward information
flows \cite{sab2009}.  Li and Zdancewic \cite{li2005} proposed an
expressive mechanism called {\sl relaxed noninterference} for
declassification policies that supports the extensional specification
of secrets and their intended declassification. A declassification
policy is a function that captures the precise information on a
confidential value that can be {\sl declassified}.  For the password
checker example,  the following declassification policy 
$\lambda p.\lambda x.h(p)==x$, allows an equality comparison with the
hash of password to be declassified (and made public), but disallows
arbitrary declassifications such as revealing the password.

The problem of information-flow security has been studied also for low
level languages.  Barthe and Rezk \cite{bar2003,bar2007} provide a
flow sensitive type system for a sequential bytecode language.  As it is
the case for most analyses, implicit flows are forbidden, and hence,
modifications of parts of the environment with lower security type than
the current context are not allowed. Genaim and Spoto present in
\cite{gen2005} a compositional information flow analysis for full Java
bytecode.
\paragraph{{\bf Information flow applied to detecting side-channel
leakages.}}  Information-flow analyses track the flow of information
through the program but often ignore information flows through side
channels. Side-channel attacks
extract sensitive information about a program's state through its
observable use of resources such as time or memory.  Several
approaches in language-based security use security type systems to
detect timing side-channels \cite{aga2000,hed2005}.  Agat  \cite{aga2000}
presents a type system sensitive to timing for a small While-language which
includes a transformation which takes a program and transforms it
into an equivalent program without timing leaks.
Molnar et al \cite{mol2005} introduce the program counter model, which
is equivalent to path non-interference, and give a program
transformation for making programs secure in this model. 

FlowTracker \cite{rodrigues2016} allows to statically
detect time-based side-channels in LLVM programs. Relying on the
assumption that LLVM code is in SSA form, they
compute control dependencies using a sparse
analysis~\cite{Choi1991} without building the whole Program Dependency
Graph. 
Leakage at assembly-level is also considered in \cite{bar2014}.  They
propose a fine-grained information-flow analysis for checking that
assembly programs generated by CompCert are constant-time. 
Moreover, they consider a stronger adversary which controls the scheduler and the cache.  

All the above works do not consider publicly observable outputs.  The
work that is closest to ours is \cite{almeida2016}, where the authors
develop a formal model for constant-time programming policies.  The
novelty of their approach is that it is distinguishing not only
between public and private input values, but also between private and
publicly observable output values. As they state, this distinction
poses interesting technical and theoretical challenges.
Moreover, constant-time implementations in cryptographic libraries
like OpenSSL include optimizations for which paths and addresses can
depend not only on public input values, but also on publicly
observable output values. Considering only input values as non-secret
information would thus incorrectly characterize those implementations
as non-constant-time. They also develop a verification technique based
on the self-composition based approach \cite{bar2004}. They reduce the
constant time security of a program P to safety of a product program Q
that simulates two parallel executions of P. The tool operates at the
LLVM bytecode level. The obtained bytecode program is transformed into
a product program which is verified by the Boogie verifier
\cite{bar2005} and its SMT tool suite. Their approach is complete only
if the public output is ignored. Otherwise, their construction relies
on identifying the branches whose conditions can only be declared
benign when public outputs are considered. For all such branches, the
verifier needs to consider separate paths for the two simulated
executions, rather than a single synchronized path and in the worst
case this can deteriorate to an expensive product construction.

\section{Conclusion and Perspectives}\label{concl:sec}
\label{sec:conclusion}

In this paper we proposed a static approach to check if a program is
output-sensitive constant-time, i.e., if the leakage induced through
branchings and/or memory accesses do not overcome the information
produced by (regular) observable outputs. 
Our verification technique is based on a so-called output-sensitive
non-interference property, allowing to compute the dependencies of a
leakage variable from both the initial values of the program inputs and
the final values of its outputs. We developed a type system on a
high-level {\bf While} language, and we proved its soundness. Then we
lifted this type system to a basic LLVM-IR and we developed a prototype tool 
operating on this intermediate representation, showing the applicability of our technique.

This work could be continued in several directions. One limitation of
our method arising in practice is that even if the two snippets
$x_l=h;o=h$ and $o=h;x_l=o$ are equivalent, only the latter can be
typed by our typing system. We are currently extending our approach by
considering also an under-approximation $\beta(\bullet)$ of the
dependencies between variables and using ``symbolic dependencies'' also for non-output variables.
Then the safety condition from Theorem \ref{l:sec} can be improved to  something like ``$\exists V$ such that $(\Gamma'(x_l)\lhd_\alpha V) \sbs (\Gamma(X_I)\lhd_\alpha V) \js (\beta'(X_O) \lhd_\alpha V) \js \alpha(X_O)$''. In the above example, we would obtain
$\Gamma'(x_l) = \alpha(h) =  \beta'(o) \sbs \alpha(o) \js \beta'(o)$, meaning that the
unwanted maximal leakage $\Gamma'(x_l)$ is less than the minimal leakage $\beta'(o)$ due to the normal output.
From the implementation point of view, further developments are needed in order to extend our prototype to a complete tool able to deal with real-life case studies.
This may require to refine our notion of arrays and to take into
account arrays and pointers as output variables. 
We could also consider applying a sparse analysis, as in FlowTracker~\cite{rodrigues2016}.  
It may happen that such a pure static analysis would be
too strict, rejecting too much ``correct'' implementations.
To solve this issue, a solution would be to combine it with the dynamic 
verification technique proposed in~\cite{almeida2016}.  Thus, our analysis could be used 
to find automatically which branching conditions are benign in the output-sensitive sense, 
which could reduce the product construction of~\cite{almeida2016}.  Finally, 
another interesting direction would be to adapt our work in the context of quantitative analysis for program leakage, like in~\cite{CacheAudit}.

\section*{Acknowledgment}
  \noindent The authors wish to acknowledge the anonymous referees for their fruitful comments and suggestions.





%
%
%
\bibliographystyle{splncs04}
\bibliography{cstTime}
\end{document}